\documentclass[final,3p,times,twocolumn]{elsarticle}
\usepackage{graphicx}
\usepackage{caption}
\usepackage{subcaption}
\usepackage[usenames,dvipsnames,svgnames,table]{xcolor}
\usepackage{multirow}
\usepackage{color}

\usepackage{amssymb}
\usepackage{amsmath}
\usepackage{algorithm}
\usepackage{algorithmic}
\usepackage{placeins}
\renewcommand{\vec}[1]{\mathbf{#1}}
\newcommand{\be}{\begin{equation}}
\newcommand{\ee}{\end{equation}}
\newcommand{\bea}{\begin{eqnarray}}
\newcommand{\eea}{\end{eqnarray}}
\usepackage{lineno}

\journal{Computer Physics Communications}

\begin{document}

\begin{frontmatter}



\title{Memory transfer optimization for a lattice Boltzmann solver on Kepler architecture nVidia GPUs}


\author{Mark J. Mawson\corref{MJM}}

\cortext[MJM]{mark.mawson@postgrad.manchester.ac.uk}
\author{Alistair J. Revell}

\address{University of Manchester, Oxford Road, Manchester, Greater Manchester, M13 9PL, UK}

\begin{abstract}
The Lattice Boltzmann method (LBM) for solving fluid flow is naturally well suited to an efficient implementation for massively parallel computing, due to the prevalence of local operations in the algorithm. This paper presents and analyses the performance of a 3D lattice Boltzmann solver, optimized for third generation nVidia GPU hardware, also known as `Kepler'. We provide a review of previous optimisation strategies and analyse data read/write times for different memory types. In LBM, the time propagation step (known as streaming), involves shifting data to adjacent locations and is central to parallel performance; here we examine three approaches which make use of different hardware options. Two of which make use of `performance enhancing' features of the GPU; shared memory and the new \textit{shuffle} instruction found in Kepler based GPUs. These are compared to a standard transfer of data which relies instead on optimised storage to increase coalesced access. It is shown that the more simple approach is most efficient; since the need for large numbers of registers per thread in LBM limits the block size and thus the efficiency of these special features is reduced. Detailed results are obtained for a D3Q19 LBM solver, which is benchmarked on nVidia K5000M and K20C GPUs. In the latter case the use of a read-only data cache is explored, and peak performance of over 1036 Million Lattice Updates Per Second (MLUPS) is achieved. The appearance of a periodic bottleneck in the solver performance is also reported, believed to be hardware related; spikes in iteration-time occur with a frequency of around 11Hz for both GPUs, independent of the size of the problem.

\end{abstract}


\end{frontmatter}

% \linenumbers

\section{ Introduction} 
The implementation of Lattice Boltzmann method (LBM) solvers on Graphics Processing Units (GPUs) is becoming increasingly popular due to the intrinsic parallelizable nature of the algorithm. A growing literature exists in this area, though with frequent hardware changes there is a need to constantly review the means of obtaining optimal performance. As a derivative of  Lattice Gas Cellular Automata (LGCA), LBM defines macroscopic flow as the collective behaviour of underlying microscopic interactions \cite{Chen1998}. The LBM improves upon LGCA by describing each point in space using a mesoscopic particle distribution function rather than an individual particle, reducing statistical noise within the method. LBM has been used extensively  in literature over the past decade and is now regarded as a powerful and efficient alternative to the classical Navier-Stokes solvers (see \cite{Chen1998,Succi2001} for a complete overview). Despite early suggestions in the literature to the contrary, Shan et al. \cite{Shan2006} formally demonstrated how LBM can return exact Navier-Stokes even for high Reynolds and high Mach number flows.

Algorithmically, the method consists of a local relaxation  (`collide') and a linear advection (`stream') of a number of discrete components of the distribution function, rendering the method highly parallelizable. Implementation of the LBM on GPUs can be traced back to 2003, when Li et al. \citep{Li2003} mapped a 2D LBM algorithm to texture and rasterization operations within the graphics pipeline. Since then, a variety of both two and three dimensional models have been implemented and GPU based LBM algorithms have been proposed for a range of applications; e.g. free surface \cite{Janßen2011}, thermal \cite{Obrecht2011}, biomedical \cite{Miki2012}.

Another early attempt was made by Ryoo et al. \cite{Ryoo2008}, who tested a CUDA port of a simple LBM code from the SPEC CPU2006 benchmark \cite{Henning2006}. This was followed by a more in-depth optimization reported by  T\"{o}lke and Krafczyk \cite{Tolke2008}, who implemented a 3D model with a reduced 13 component distribution function on G80 generation GPUs, specifically designed to increase maximum throughput. They used shared memory to avoid costly misaligned access to the RAM of the GPU when performing advection of the distribution function (although in general, using less components of $f_{i}$ will reduce accuracy). A more complex split propagation method was proposed by \cite{Tolke2008a}, in which the data is first shifted along the contiguous dimension within shared memory for each block, before the perpendicular shifts are performed in global memory\footnote{This paper also  provides a useful introduction to LBM implementation in CUDA.}. This approach demonstrated high levels of efficiency, but necessitated careful handling of data entering/leaving each block. Habich et al. \cite{Habich2011} extended this work to the D3Q19 model (the same model presented in this work).

Obrecht et al. \cite{Obrecht2011c} identified that, contrary to previous attempts to avoid misalignment at all cost, the cost of a misaligned access in shared memory was actually similar to that caused by a global memory exchange; thus they proceeded to investigate the potential for performance improvement brought about by avoiding the use of shared memory altogether. Indeed, previous advances in nVidia hardware, first from compute capability 1.0 to 1.3, and then on to 2.0, substantially improved the efficiency of misaligned memory transactions; this had the important consequence that the use of shared memory was no longer so crucial. Furthermore,  \cite{Obrecht2011a} reported the cost of a misaligned read to be less than a misaligned write; an observation leads one to prefer the `pull' algorithm to the `push', as discussed in Section \ref{sect:impl}. A peak performance of 516 Million Lattice Updates Per Second (MLUPS)\footnote{This is a common performance measure based on the number of lattice points that can be updated every second.} was reported with a maximum throughput of 86\%. In addition, it is noted that previous works using Shared Memory were highly optimised, and thus obtaining substantial extra performance would not be trivial. More importantly, the implementation of more complex code to handle e.g. multiple distribution functions or extra body force terms would only be possible with a significant reduction of performance. Indeed, the present work is building towards an efficient GPU implementation of the Immersed Boundary Method with LBM reported in \cite{Favier2013}, and thus this reasoning is of high relevance to our work (see  \cite{Mawson2013}).

A comprehensive series of further work by Obrecht et al. \cite{Obrecht2011c,  Obrecht2013, obrecht2013b}  focused on compute 2.x capable hardware in their development of a multi-GPU implementation of a Hybrid thermal LBM scheme based on D3Q19, and did not make use of shared memory. Later, Habich et al. \cite{Habich2012}, also presented implementations for `Fermi' generation GPUs  without the use of shared memory. 

%These works provided first evidence that improved memory performance for misaligned accesses effectively rendered the use of shared memory redundant for block sizes typical to LBM algorithms.

In the present work, particular attention is paid to a comparison of three methods of performing the advection operation; the first which performs the propagation directly in the GPUs RAM (DRAM), a second that utilizes a shared memory space as an intermediate buffer and a third, new, method that performs the propagation locally within a group of 32 threads without using any intermediate memory using the \textit{shuffle} instruction new to GPUs based on the Kepler architecture\cite{NVIDIACorporation2012}. The results of this comparison are then used to implement an efficient 3D LBM solver on first and second generation Kepler GPUs (compute capabilities 3 and 3.5 respectively). In the case of second generation Kepler GPUs a read-only cache is also enabled to provide a small, but measurable, improvement in performance.
In the following sections the architecture and programming model for CUDA based GPUs is introduced, before the mathematical description and specific form of the LBM used is given. Analysis of the three propagation techniques is then performed, along with consideration of other implementation aspects and estimation of the maximum achievable performance of a LBM solver based on memory requirements.

	\section{GPU Computing}
	
\begin{figure}[t!]
\begin{center}
\includegraphics[width=\linewidth]{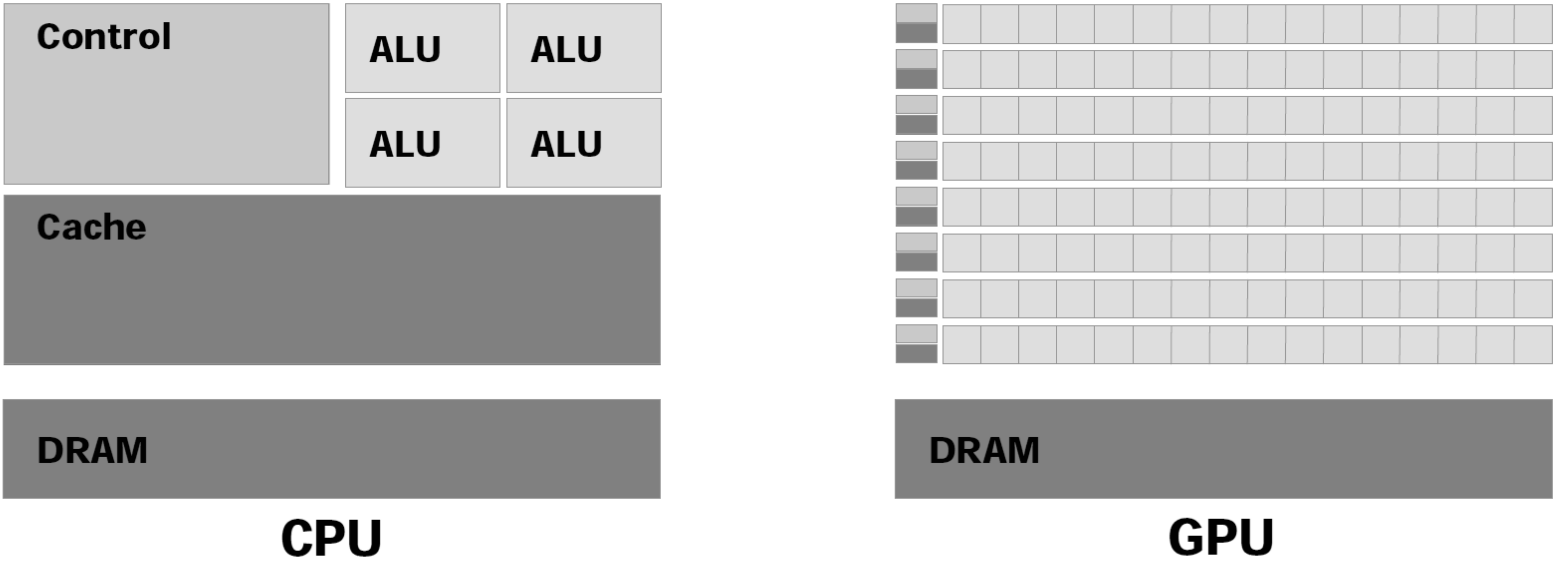}     
\end{center}
\vspace{-10pt}
\caption{Typical GPU and CPU architectures \cite{NVIDIA2009a}}\label{fig:GPUvsCPU}
\end{figure}	

\begin{figure}[t!]
\begin{center}
\includegraphics[width=0.9\linewidth]{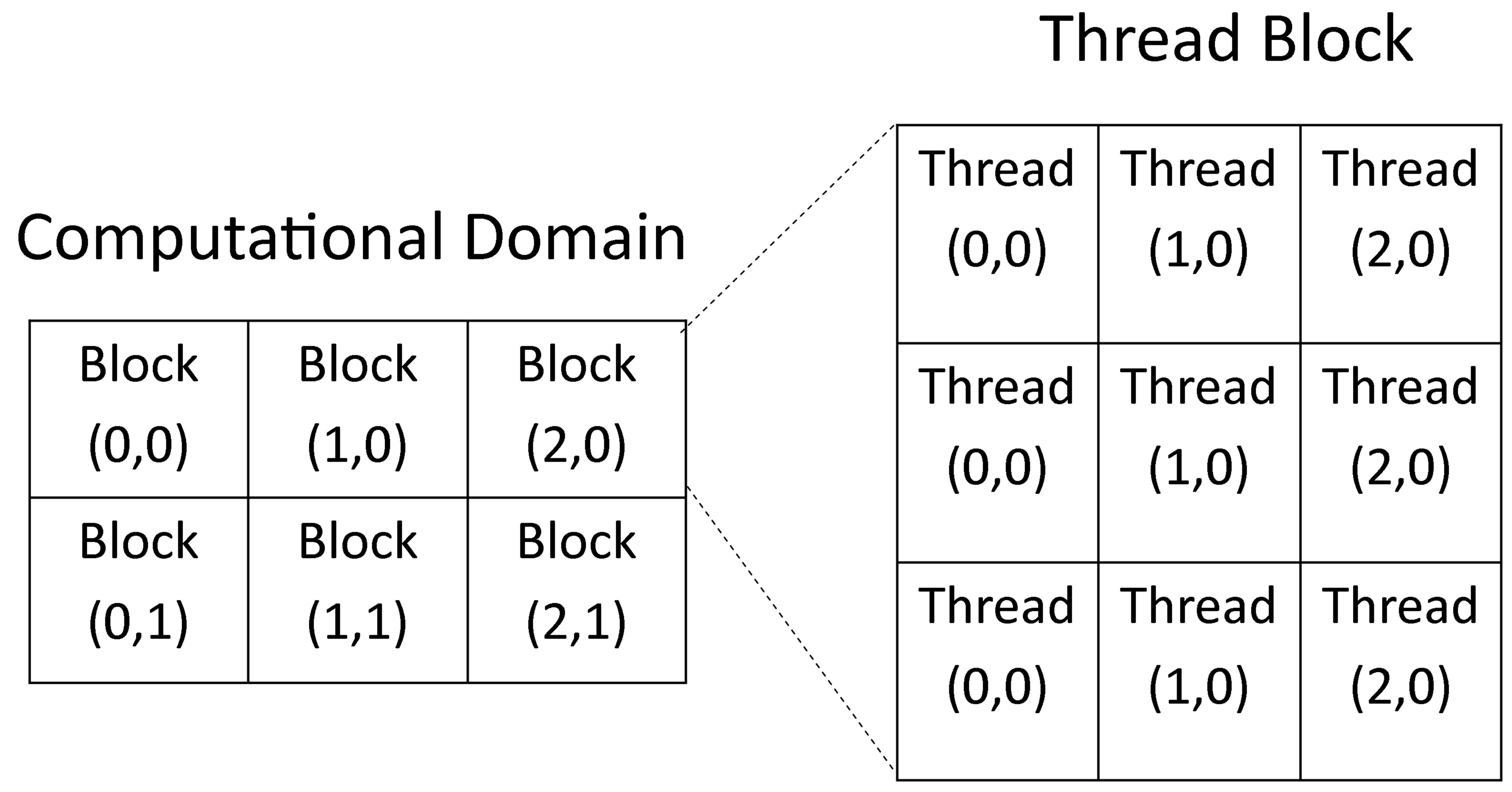}     
\end{center}
\vspace{-10pt}
\caption{Blocks \& Threads in CUDA  \cite{NVIDIAPG3.2}.}\label{fig:BlocksThreads}
\vspace{-10pt}
\end{figure}

\begin{figure*}[t!]
\begin{center}
\includegraphics[width=0.8\linewidth]{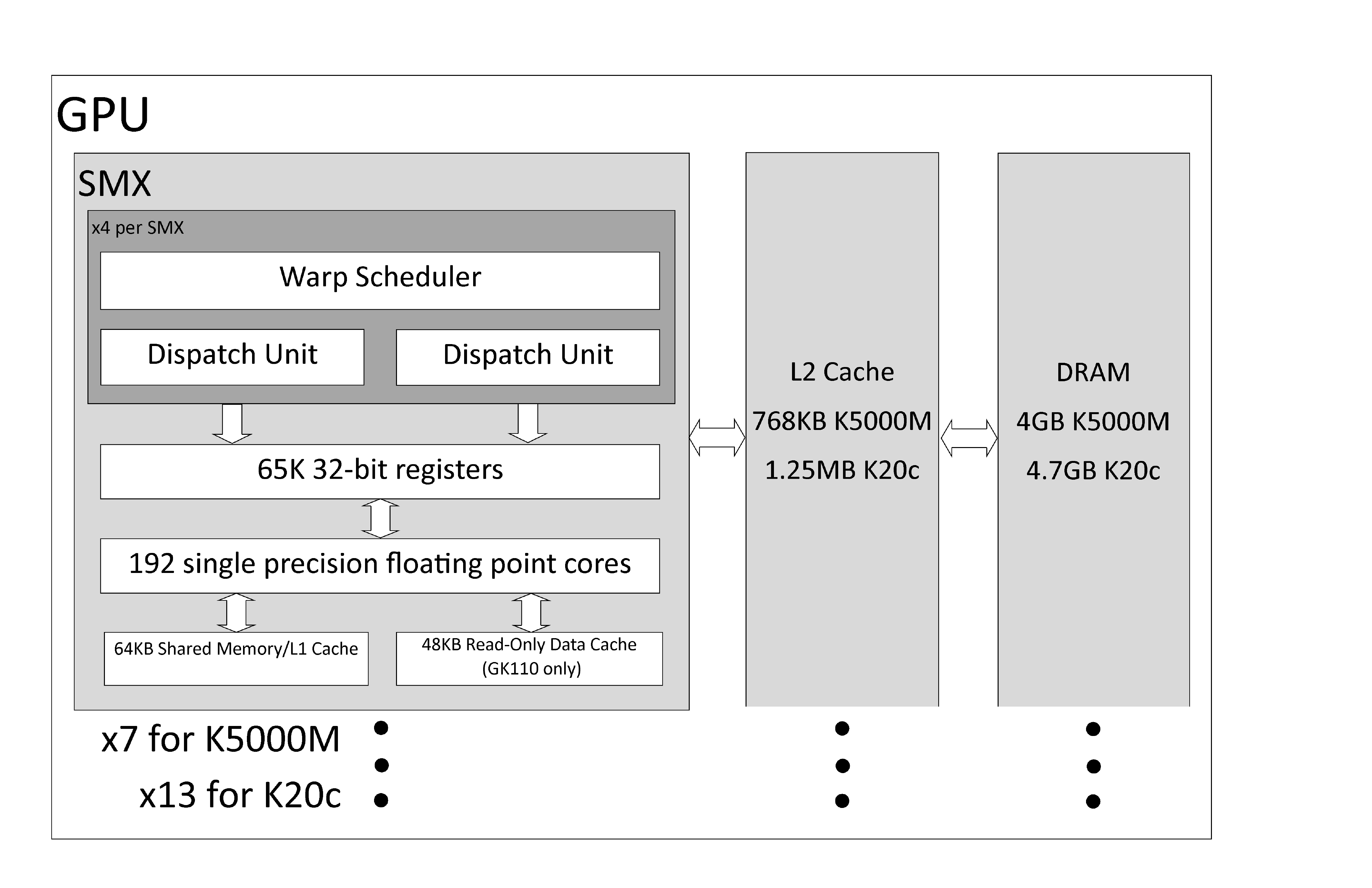}     
\end{center}
\vspace{-10pt}
\caption{Typical Kepler GPU Architecture}\label{fig:KeplerSMX}
\vspace{-10pt}
\end{figure*}		
	
	Modern GPUs use a Unified Shader Architecture \cite{Luebke2007}, in which blocks of processing cores (capable of performing operations within all parts of the graphics pipeline) are favored over hardware in which the architecture matches the flow of the graphics pipeline. For the purpose of this paper it is sufficient to note that the graphics pipeline takes the vertices of triangles as inputs and performs several vertex operations, such as spatial transformations and the application of a lighting model, to create a `scene'. The scene is then rasterized to create a 2D image which can then have textures placed over the component pixels to create the final image. For a more comprehensive introduction to the graphics pipeline and how older GPUs matched their architecture to it see \cite{Pharr2005}.

	The architecture of a generic Unified Shader based GPU is shown alongside that of a generic CPU in Figure 1. Processing cores can be seen arranged into rows with small amounts of cache and control hardware; the combination of all three is known as a Streaming Multiprocessor (SMX). Comparison with a generic CPU highlights the following key differences:  
	
	\begin{enumerate}
	\item  GPUs sacrifice larger amounts of cache and control units for a greater number of processing cores.
	
	\item  The cores of a SMX take up less die space than those of a CPU, and as a result are required to be more simple in design.
	\end{enumerate}
	
	These attributes render the GPU suitable for performing computation on large datasets where little control is needed, i.e. the same task is performed across the entire dataset. Indeed, it is this aspect which has generated interest in GPU computing for the Lattice Boltzmann Method, as identical independent operations are performed across the majority of the fluid domain, boundary conditions being the obvious exception.

	\subsection{ Threads and Blocks of Processing}
	
	Code written for nVidia GPUs is generally parallelized at two levels; the computation is divided into blocks which contain component parallel threads (see Figure 2). A single block of threads is allocated to a SMX at any one time, with the component threads divided into groups of 32 called `warps'. The threads within a warp are executed in parallel, and all threads within the warp must execute the same instruction or stall. Figure 2 shows an example of threads and blocks being allocated two-dimensionally; the division of threads and blocks can be performed in \textit{n} dimensions, depending on the problem.

\subsection{nVidia Kepler Architecture GPUs}\label{sec:Kepler}
	In this paper two Kepler architecture GPUs are tested; the K5000M and the K20c. Details of the general hardware layout for both devices is provided in Figure \ref{fig:KeplerSMX}.  The K5000M is based on the first generation GK104 Kepler architecture, and as such is limited in it's double precision performance, with a ratio of 24:1 between single and double precision peak performance. Processing cores are grouped into blocks of 192, known as Streaming Multiprocessors (SMX), in order to share access to 64KB of configurable cache/shared memory, and four instruction schedulers capable of dispatching two instructions per clock cycle\footnote{Provided the instructions are from the same warp and independent in nature.}. Seven SMXs make up the K5000M, and these SMXs share access to 512KB of L2 cache and 4GB of DRAM. In the configuration used in this paper, the K5000M is clocked at 601MHz, giving a theoretical peak single precision performance of 1.62 TFLOPS. Peak DRAM bandwidth was measured as 64.96GB/sec using the benchmarking program included in the CUDA SDK. 
	
	The K20c is based on the newer GK110 architecutre, which is largely the same as a GK104, the most significant difference being extra double precision units within each SMX to improve the single/double precision performance ratio to 3:1. A K20c contains 13 SMXs, which share access to 1.25MB of L2 cache and 4.7GB of DRAM. The theoretical peak single precision performance of the configuration used in this work is 3.5 TFLOPS, and measured bandwidth was 157.89GB/sec.

\section{ The Lattice Boltzmann Method}

\subsection{ Numerical method}

In this study the Lattice Boltzmann Method is used to simulate fluid flow, this method is based on microscopic models and mesoscopic kinetic equations; in contrast to Navier-Stokes which is in terms of macroscale variables. The Boltzmann equation for the probability distribution function $f=f(\vec{x},\vec{e},t)$ is given as follows:
\begin{equation}\label{eqn:Boltzmann}
\frac{\partial f}{\partial t} +\vec{e}\cdot \nabla _{\vec{x}} f=\Omega
\end{equation}
where $x$ are the space coordinates, and $e$ is the particle velocity. The collision operator $\Omega$ is simplified using the `BGK' single time relaxation approach found in \cite{Bhatnagar1954}, in which context, it is assumed that local particle distributions relax to an equilibrium state, $f^{(eq)} $in time $\tau$:
\begin{equation}\label{eqn:LBGK}
\Omega =\frac{1}{\tau } \left(f^{(eq)} -f\right)
\end{equation}

\begin{figure}[h!]
		\begin{center}
		 \includegraphics[width=0.6\linewidth]{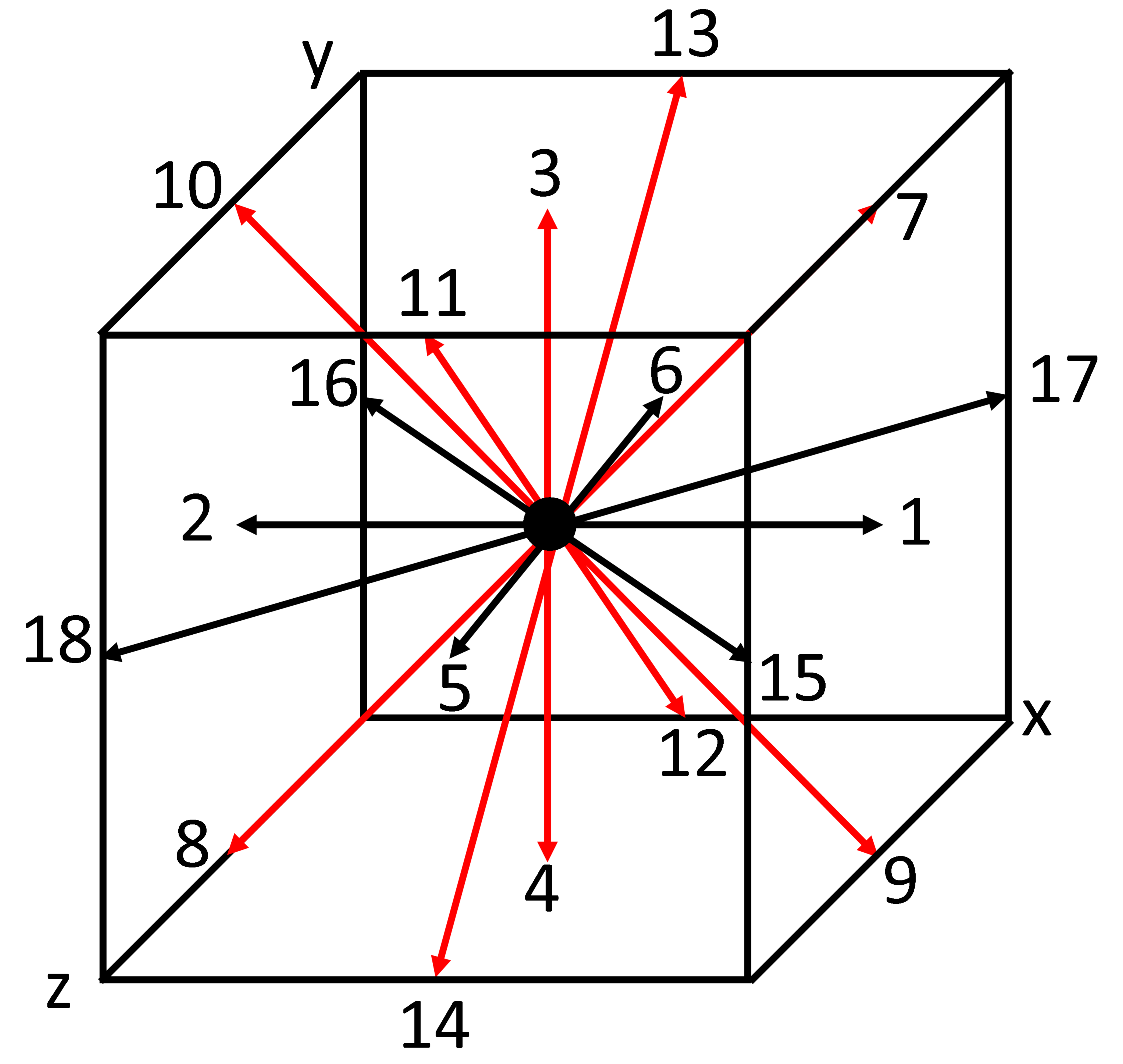}     
		\end{center}
	\vspace{-10pt}
		\caption{The D3Q19 Lattice.}\label{fig:D3Q19}
%	\vspace{-10pt}
	\end{figure}

The discretized form of Eqn. \ref{eqn:Boltzmann} is obtained via Taylor series expansion following \cite{He1997b}, as shown in Eqn. \ref{eqn:DiscreteLBGK}, where $f_i$ refers to the discrete directions of $f$. The dimensionality of the model and spatial discretization of $f$  is given in the `D\textit{m}Q\textit{n}' format, in which the lattice has \textit{m} dimensions and $f$ has \textit{n} discrete directional components. In the current work the  D3Q19  model is used, in which the discrete velocity is defined according to Eqn. \ref{eqn:D3Q19ei} and is visualized in Figure \ref{fig:D3Q19}. Since spatial and temporal discretization in the lattice are set to unity, the lattice speed $c=\Delta x/\Delta t=1$. 

\begin{equation}
f_i(\vec{x}+\vec{e}_i\Delta t, t+\Delta t)=f_i(\vec{x},t)+\frac{1}{\tau}\left[f_i^{eq}(\vec{x},t)-f_i(\vec{x},t)\right]\label{eqn:DiscreteLBGK}
\end{equation}

\begin{footnotesize}

\setcounter{MaxMatrixCols}{19}
\setlength{\arraycolsep}{2pt}
\begin{align}\label{eqn:D3Q19ei}
e_{i} =c\begin{pmatrix} 0 & 1 & -1 & 0 & 0 & 0 & 0 & 1 & -1 & 1 & -1 & 0 & 0 & 0 & 0 & 1 & -1 & -1 & 1 \\ 0 & 0 & 0 & 1 & -1 & 0 & 0 & 1 & -1 & -1 & 1 & 1 & -1 & 1 & -1 & 0 & 0 & 0 & 0 \\ 0 & 0 & 0 & 0 & 0 & 1 & -1 & 0 & 0 & 0 & 0 & 1 & -1 & -1 & 1 & 1 & -1 & 1 & -1 \end{pmatrix}
 \notag \\(i=0,1,...,19)
 \end{align}
 
\end{footnotesize}
 
The equilibrium function $f^{(eq)} \left(x,t\right)$ can be obtained by Taylor series expansion of the Maxwell-Boltzmann equilibrium distribution \cite{Qian1992} :
\begin{equation}\label{eqn:LBGKeq}
f_{i}^{eq} =\rho \omega _{i} \left[1+\frac{\vec{e}_{i} \cdot \vec{u}}{c_{s}^{2} } +\frac{\left(\vec{e}_{i} \cdot \vec{u}\right)^{2} }{2c_{s}^{4} } -\frac{\vec{u}^{2} }{2c_{s}^{2} } \right] 
\end{equation}

In Eqn. \ref{eqn:LBGKeq}, $c_{s}$ is the speed of sound $c_{s} =1/\sqrt{3} $ and the coefficients of $\omega _{i}$ are $\omega _{0} =1/3$, $\omega _{i} =1/18,\; i=1..6$ and $\omega _{i} =1/36,\; i=7..19$ . %The macroscopic velocity $\textbf{U}$(made up of $\textit{u, v}$ and $\textit{w})$  must satisfy the requirement for low Mach number, $M$, i.e. that$|u|/c_{s} \approx M<<1$.

Macroscopic quantities (moments of the distribution function) are obtained as follows:
\begin{equation}\label{eqn:rhof}
\rho =\mathop{\sum  }\limits_{i} f_{i} 
\end{equation}
\begin{equation}\label{eqn:uf}
\rho \vec{u}=\mathop{\sum  }\limits_{i} e_{i} f_{i} 
\end{equation}
The multi-scale expansion of Eqn. \ref{eqn:DiscreteLBGK} neglecting terms of \textbf{O}($\varepsilon $\textit{$M^2$}) and using Eqns. \ref{eqn:rhof} and \ref{eqn:uf} returns the Navier-Stokes equations to second order accuracy \cite{He1997b}, allowing the LBM to be used for fluid simulations.

\section{Implementation}\label{sect:impl}
%This section describes alterations made to the Lattice Boltzmann Method in order to expose more parallelism and reduce synchronization, as well as the basic memory requirements of the model and it's efficient implementation. Finally a comparison of three different models used to propagate values of $f$ during the streaming stage of the LBM is performed.

\subsection{Memory Arrangement}\label{sec:LBMMemory}

The present solver is parallelized such that one thread will perform the complete LBM algorithm at one spatial location $f\left(x\right)$ in the fluid domain. Each thread stores values of $f\left(x\right)$, $\rho$, \textbf{u} and information about whether or not the current location is a boundary in a struct within register space to minimize high latency access to DRAM once initially loaded. Within DRAM it is common practice to `flatten' multiple dimension arrays into a single dimension, as the extra de-referencing operations required add to the already large latency of accessing off-chip memory. This is extended to combining and flattening the components of \textit{f} into a single array, such that $f_{i} \left(x\right)$ is addressed as f[i*Nx*Ny*Nz+z*Ny*Nx+y*Nx+x]. Storing \textit{f} in order of \textit{i} and then by spatial coordinates will cause neighbouring threads within a warp to access contiguous memory in \textit{f}. If these accesses are aligned within a 128-byte segment (see Section \ref{sec:StreamAccess}) the data transactions can be grouped into a single larger transaction; i.e. resulting in a coalesced access.

\begin{figure}[h!]
	\begin{center}
 \includegraphics[width=0.9\linewidth]{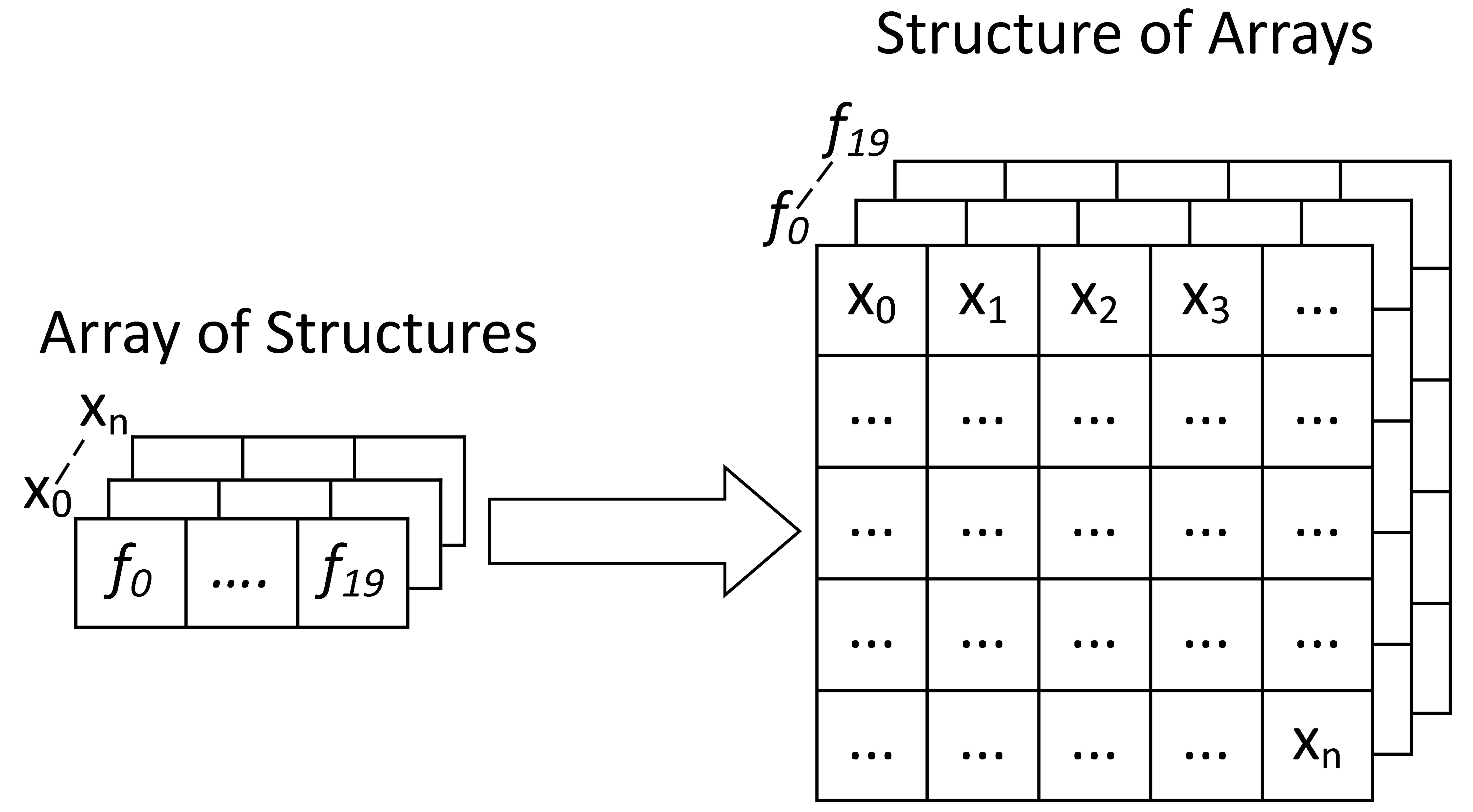}     
	\end{center}
	\vspace{-10pt}
	\caption{Arrays of Structures and Structures of Arrays.}\label{fig:SOAs}
\end{figure}

The ordering of the data in $f$ is known as storing in a `Structure of Arrays' (SoA) format; without the collapsing of the different components of $f$ into a single array this can be seen as a structure containing 19 arrays, each one corresponding to all of the spatial locations of one component of $f$. The opposite `Array of Structures' (AoS) arrangement is shown in Figure \ref{fig:SOAs} for clarity, which corresponds to an a single array with one element per spatial location, each containing a data structure to store the 19 components of $f$. Once $f$ has been read into a core, an analogy can be drawn between this format and the `array' of GPU cores, each containing their own local structure. While AoS is shown to be preferable for serial CPU implementations \cite{Pohl2003a}, SoA is necessary to improve coalesced access to global memory with GPU versions.

Without the presence of macroscopic values, a single point in the lattice requires 19 loads from global memory, and 19 stores back to global memory during an iteration of the LBM algorithm \footnote{The stores are actually performed on a redundant copy of $f$ to ensure data dependency is not violated, at the end of each timestep the pointers to the original and redundant copy of $f$ are swapped.}. In single precision this yields 152 bytes of data to be transferred for each lattice point. Using the measured bandwidth of 65GB/sec for the K5000M from Section \ref{sec:Kepler}, the theoretical limit is 459 MLUPS. For the K20c this limit is 1115 MLUPS. If macroscopic values are required, then an extra four storage operations are needed and the maximum theoretical performance drops to 415 MLUPS (1009 MLUPS for the K20c).

\subsection{Independent LBM algorithm}\label{sec:lbmAlg}

The conventional LBM is typically broken down into several steps as described in Algorithm \ref{alg:LBMConventional}, which is typically the form taken for CPU implementations. This algorithm poses some locality problems if it is to be used in a highly parallel fashion. If a thread is launched for each location $f\left(x\right)$ then the non-local operation $f_{i} \left(x+e_{i} \Delta t,t+\Delta t\right)=f_{i} \left(x,t+\Delta t\right)$ will require a synchronization across the domain (as shown) before the boundary conditions, $\rho$ and \textbf{u} can be calculated.

\begin{algorithm}
\caption{The Conventional LBM}
\label{alg:LBMConventional}
\begin{algorithmic}[1]
\begin{small}
\STATE \textcolor{green}{Kernel 1}
\FORALL{Locations $x$ in $f_i\left(x,t\right)$}
\FORALL{$i$}
\STATE Read $f_{i} \left(x,t\right)$ from memory to a local store $fLocal_i$.
\ENDFOR
\STATE Calculate $f^{eq}$  using Eqn. \ref{eqn:LBGKeq}.
\STATE Perform Collision  $fLocal_{i} =fLocal_{i} \left(x,t\right)+\frac{\Delta t}{\tau } \left(f^{(eq)} \left(x,t\right)-f\left(x,t\right)\right)$.
\STATE Stream $fLocal_i$ to the location $f_{i} \left(x+e_{i} \Delta t,t+\Delta t\right)$.
\ENDFOR
\STATE \textcolor{red}{Synchronization across $f_{i} \left(x+e_{i} \Delta t,t+\Delta t\right)$}
\STATE \textcolor{blue}{Kernel 2}
\FORALL{Locations $x$ in $f_i\left(x,t\right)$}
\FORALL{$i$}
\STATE Read $f_{i} \left(x,t\right)$ from memory to a local store $fLocal_i$.
\ENDFOR
\STATE Apply boundary conditions.
\STATE Calculate $\rho$ and \textbf{u} using Eqns. \ref{eqn:rhof} and \ref{eqn:uf}.
\ENDFOR
\end{small}

\end{algorithmic}
\end{algorithm}

Instead, one of the two re-orderings presented in \cite{Wellein2006} can be used. These are described as `push' or `pull' algorithms, depending on whether the streaming operation (which causes misaligned access to DRAM, see Section \ref{sec:StreamAccess}) occurs at the end ($f_{i} \left(x+e_{i} \Delta t,t+\Delta t\right)=f_{i} \left(x,t+\Delta t\right)$) or beginning ($f_{i} \left(x,t\right)=f_{i} \left(x-e_{i} \Delta t,t-\Delta t\right)$) of the algorithm. Both algorithms remove the need for an additional synchronization by placing the synchronization point at the end of an iteration, where a synchronization implicitly occurs as the loop (or kernel when implemented in CUDA) across the domain exits. This also eliminates the requirement to store $\rho$ and \textbf{u} in DRAM, unless they are required for post-processing, as they are only used in enforcing the boundary conditions and the calculation of $f^{(eq)}$. 
\begin{algorithm}
\caption{The Push LBM Iteration}
\label{alg:LBMPush}
\begin{algorithmic}[1]
\begin{small}
\FORALL{Locations $\vec{x}$ in $f_i\left(\vec{x},t\right)$}
\FORALL{$i$}
\STATE Create a local copy of $f_i(\vec{x},t)$
\ENDFOR
\STATE Apply boundary conditions
\STATE Calculate $\rho$ and $\vec{u}$ using Eqns. \ref{eqn:rhof} and \ref{eqn:uf}.
\FORALL{$i$}
\STATE Calculate $f^{eq}$  using Eqn. \ref{eqn:LBGKeq}.
\STATE Perform Collision - $fLocal_{i} =fLocal_{i} \left(\vec{x},t\right)+\frac{\Delta t}{\tau } \left(f^{(eq)} \left(\vec{x},t\right)-f\left(x,t\right)\right)$.
\STATE Stream local copies of $f_i$ to their location \textcolor{red}{$f_(\vec{x}+\vec{e}_i,t+1)$}
\ENDFOR
\ENDFOR
\end{small}
\end{algorithmic}
\end{algorithm}

\begin{algorithm}
\caption{The Pull LBM Iteration}
\label{alg:LBMPull}
\begin{algorithmic}[1]
\begin{small}

\FORALL{Locations $x$ in $f_i\left(x,t\right)$}
\FORALL{$i$}
\STATE Stream to $fLocal_i$ from the location \textcolor{red}{$f_{i} \left(x-e_{i} \Delta t,t-\Delta t\right)$}.
\ENDFOR
\STATE Apply boundary conditions.
\STATE Calculate $\rho$ and \textbf{u} using Eqns. \ref{eqn:rhof} and \ref{eqn:uf}.
\STATE Calculate $f^{eq}$  using Eqn. \ref{eqn:LBGKeq}.
\FORALL{$i$}
\STATE Perform Collision  $fLocal_{i} =fLocal_{i} \left(x,t\right)+\frac{\Delta t}{\tau } \left(f^{(eq)} \left(x,t\right)-f\left(x,t\right)\right)$.
\ENDFOR
\ENDFOR
\end{small}

\end{algorithmic}
\end{algorithm}

\subsection{Read/Write Memory speed}
As stated in the Introduction, \cite{Obrecht2011c} examined the cost of misaligned reads and writes for compute 1.3 devices and reported that the former were more efficient than the latter; motivating their preference for the `pull' algorithm. In what follows, we provide results of a similar experiment for the more recent compute 3.0 and 3.5 devices.  aligned and misaligned read and writes to several large vectors were performed to mimic the behaviour of the `push' and `pull' algorithms, and the DRAM bandwidth achieved is measured. 

Multiple vectors are used to provide improved Instruction Level Parallelism (ILP), which is a strategy to mask some memory latency by allowing a single thread to launch several independent memory requests before previous requests have returned \cite{Volkov2010}. This helps make use of the dual-instruction dispatching feature of each warp scheduler; two instructions can only be dispatched in a single clock cycle only if they are from the same warp and the instructions are independent. In the full implementation of the LBM solver ILP is used across the various components of $f_i$ when streaming, and also when performing the collision operation. For compute 3.5 devices the read-only data cache (previously only accessible through the use of textures) is also considered.

\begin{figure*}

\begin{minipage}{0.49\textwidth}
\includegraphics[width=\linewidth]{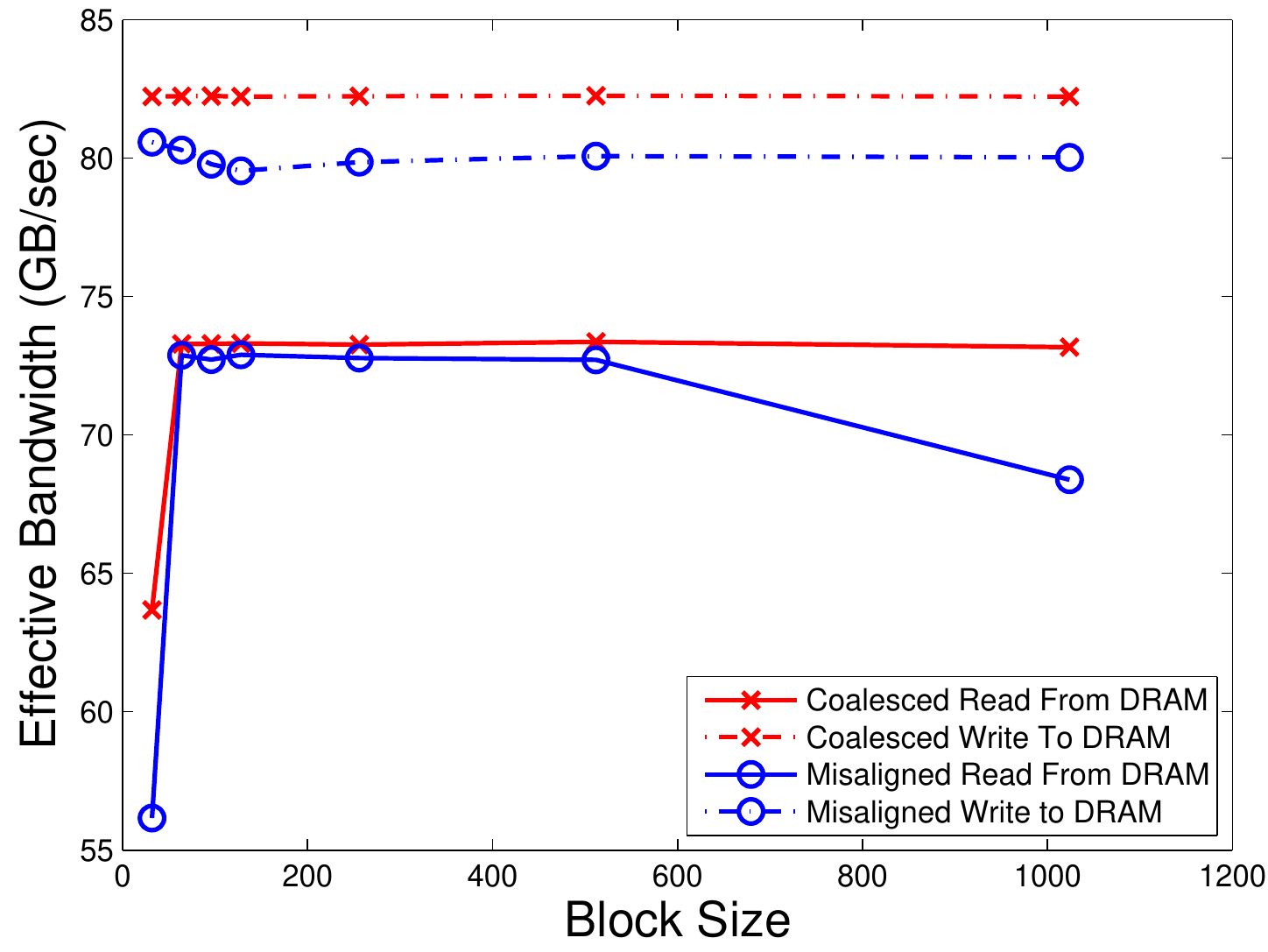}     
\vspace{-10pt}
\caption{Effective bandwidth on K5000m for aligned/misaligned reads and writes}\label{fig:ReadwriteK5000}
\vspace{-10pt}
\end{minipage}
\hspace{0.5cm}
\begin{minipage}{0.49\textwidth}
\includegraphics[width=\linewidth]{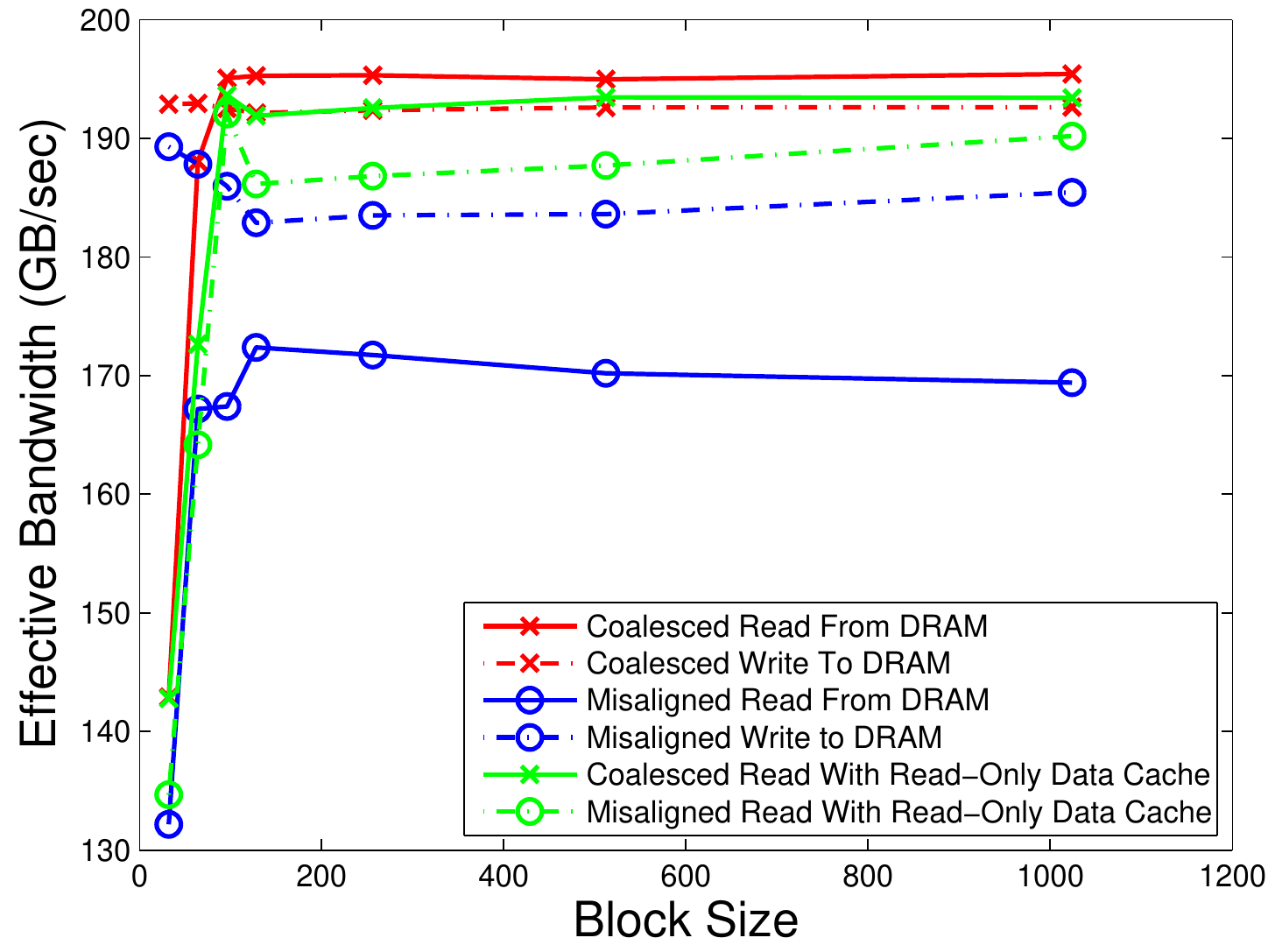}     
\vspace{-10pt}
\caption{Effective bandwidth on K20c for aligned/misaligned reads and writes}\label{fig:ReadwriteK20}
\vspace{-10pt}
\end{minipage}

\end{figure*}

Figure \ref{fig:ReadwriteK5000} provides a comparison of read and write bandwidth on the K5000m, for both coalesced and misaligned access to data. Figure \ref{fig:ReadwriteK20} displays the corresponding results for the K20c, in addition to times for Read-Only data access. Results show that misaligned reads incur a smaller penalty than misaligned writes when compared with their aligned access counterparts on the K5000m, and are therefore more efficient (99\% of aligned access bandwidth versus 96\%). On the K20c the read-only cache is required to maintain the efficiency of misaligned reads, efficiency drops to 88\% without it. The use of read-only cache does have a small detrimental effect on the performance of aligned reading from memory due to the overhead of passing through extra hardware.  Overall, it is clear that the pull algorithm is preferable. 

The re-ordered \textit{pull} algorithm is used, following \cite{Wellein2006} and \cite{Rinaldi2012} and shown in Algorithm \ref{alg:LBMPull}. The pull algorithm takes it's name from the operation $f_{i} \left(x,t\right)=f_{i} \left(x-e_{i} \Delta t,t-\Delta t\right)$, which is used instead of $f_{i} \left(x+e_{i} \Delta t,t\right)=f_{i} \left(x,t\right)$; i.e. data is loaded directly into it's new location. 

\subsubsection{Register Usage for LBM}
A single SMX in compute 3.x devices contains 65536 registers, and so the trade off between block size and grid size is best understood as follows:

	\begin{equation*}
		{\color{black}\frac{\text{registers}}{\text{thread}}}\times{\color{black}\frac{\text{threads}}{\text{block}}}\times{\color{black}\frac{\text{blocks}}{\text{SMX}}}\le 65536
	\end{equation*}

The D3Q19 solver presented in this chapter is uses approximately 45 registers/thread, depending on the boundary conditions imposed. 
 
If a block size of 1024 were used, a total of $\sim$45K threads would be needed for each block, and thus only one block could be launched, causing the remaining 20K registers to go unused.  The authors in \cite{Obrecht2013} recommend block sizes of no more than 256 threads for this reason.

\subsubsection{Access Patterns for The Streaming Operation}\label{sec:StreamAccess}
Access to global memory is most efficient when threads within a warp access data in the same 128-byte segment, when this occurs the 32 requests from each thread in the warp are coalesced into a single request. Within the streaming operation, alignment to a 128-byte segment is dependent on the value of the the $x$ component of $\vec{e}_i$. When it is zero these coalesced accesses are guaranteed, as propagation of values in the $y$ and $z$ directions move access to a different segment without any data misalignment. When the $x$ component of $\vec{e}_i$ has a non-zero value, misaligned access to the segments will occur, and two memory requests will be required per warp; one to load values from 31 addresses in the same segment and a second to load a value from the previous (if $\vec{e}_{i_x}=1$) or next (if $\vec{e}_{i_x}=-1$) segment.  

\begin{figure}[t!]
\begin{center}
 \includegraphics[width=0.9\linewidth]{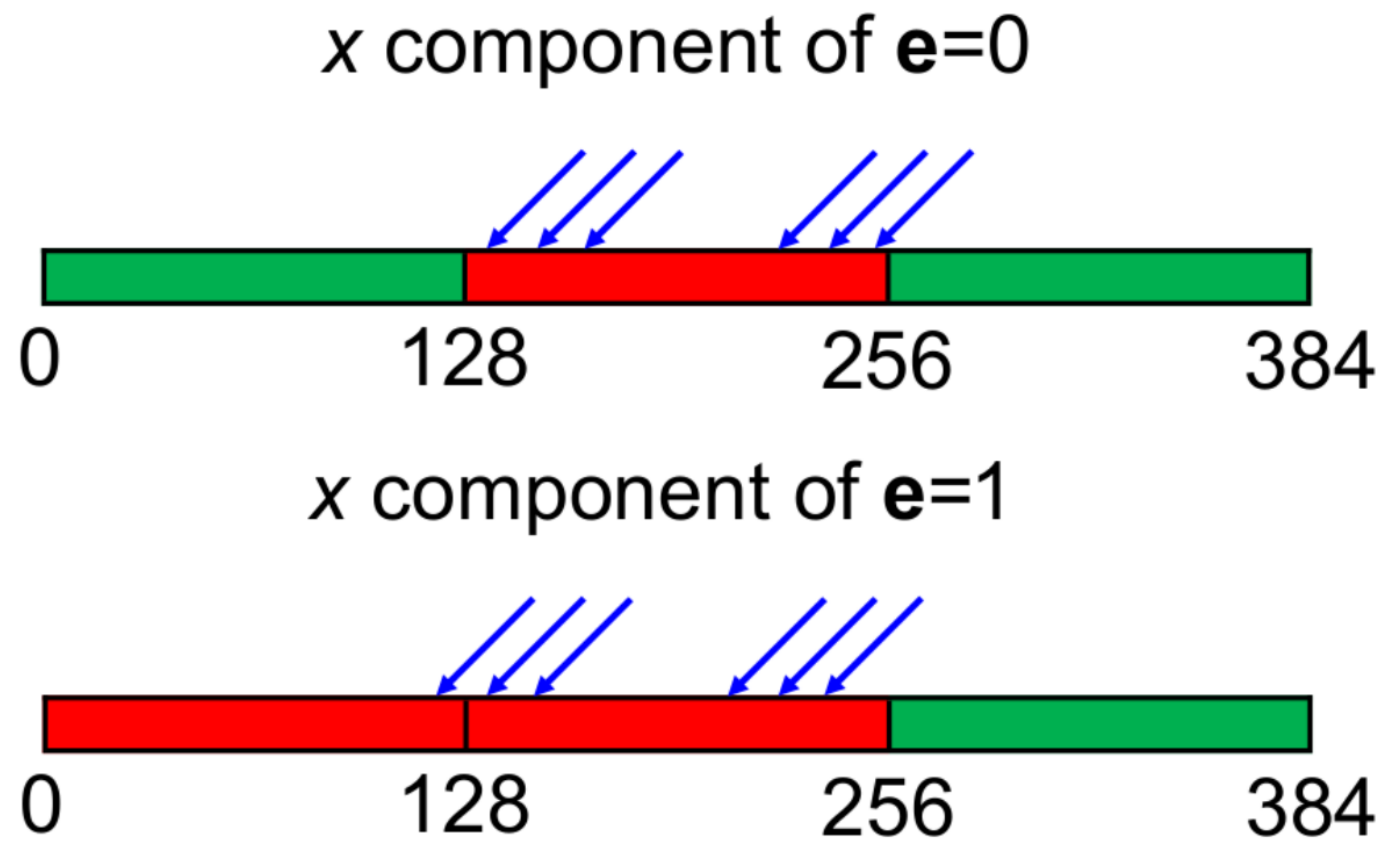}     
\end{center}
\vspace{-10pt}
\caption{Aligned and misaligned access to DRAM in the streaming operation}\label{fig:misaligned}
\vspace{10pt}

\begin{center}
\includegraphics[width=\linewidth]{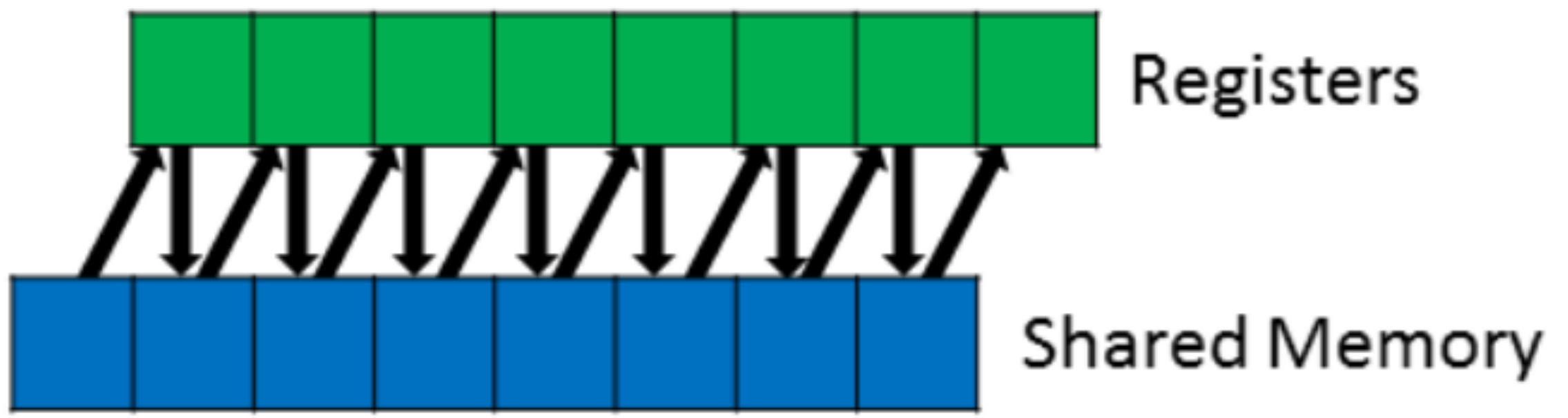}     
\end{center}
\vspace{-10pt}
\caption{Using shared memory to propagate values}\label{fig:SharedProp}
\vspace{10pt}

\begin{center}
\includegraphics[width=\linewidth]{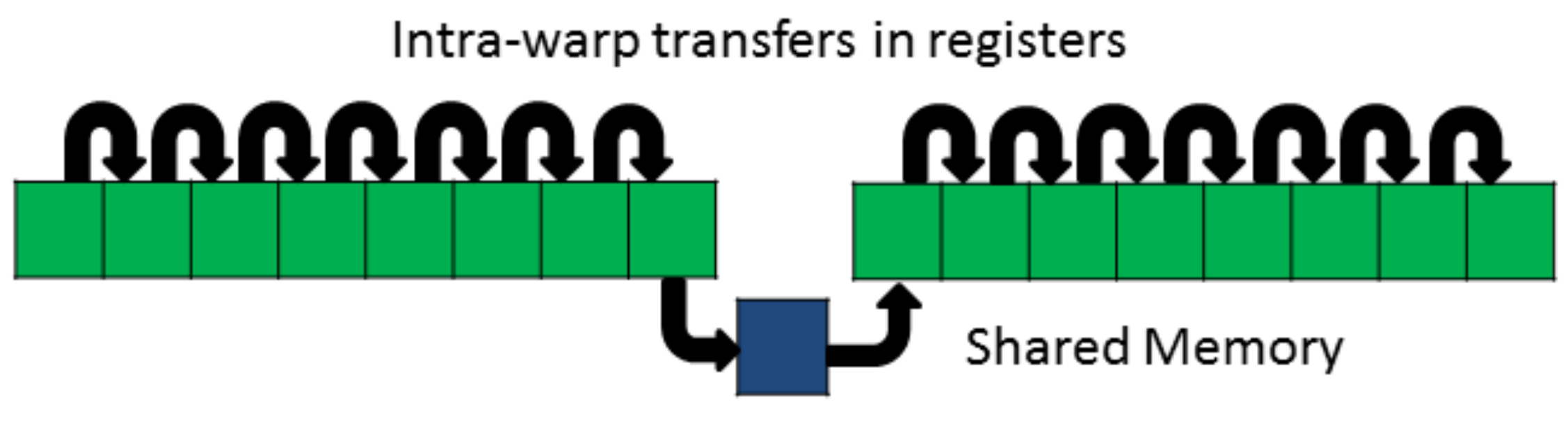}     
\end{center}
\vspace{-10pt}
\caption{Using shuffle instruction to propagate values}\label{fig:shuffleProp}
\end{figure}

\begin{figure*}

\begin{minipage}{0.49\textwidth}
\includegraphics[width=\linewidth]{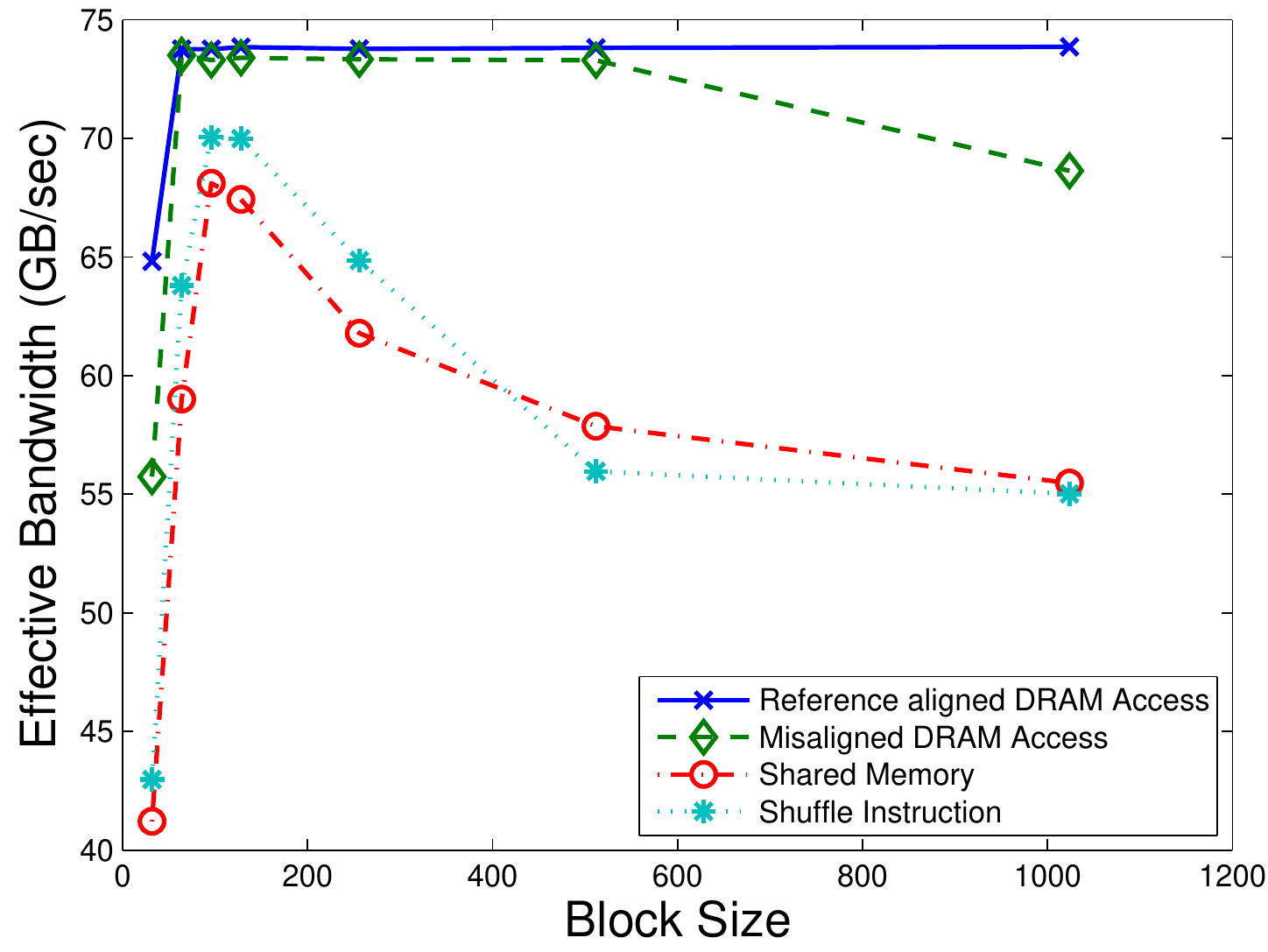}     
\vspace{-10pt}
\caption{Effective bandwidth on K5000m for offest-by-one DRAM reads}\label{fig:StreamTimesK5000}
\vspace{-10pt}
\end{minipage}
\hspace{0.5cm}
\begin{minipage}{0.49\textwidth}
\includegraphics[width=\linewidth]{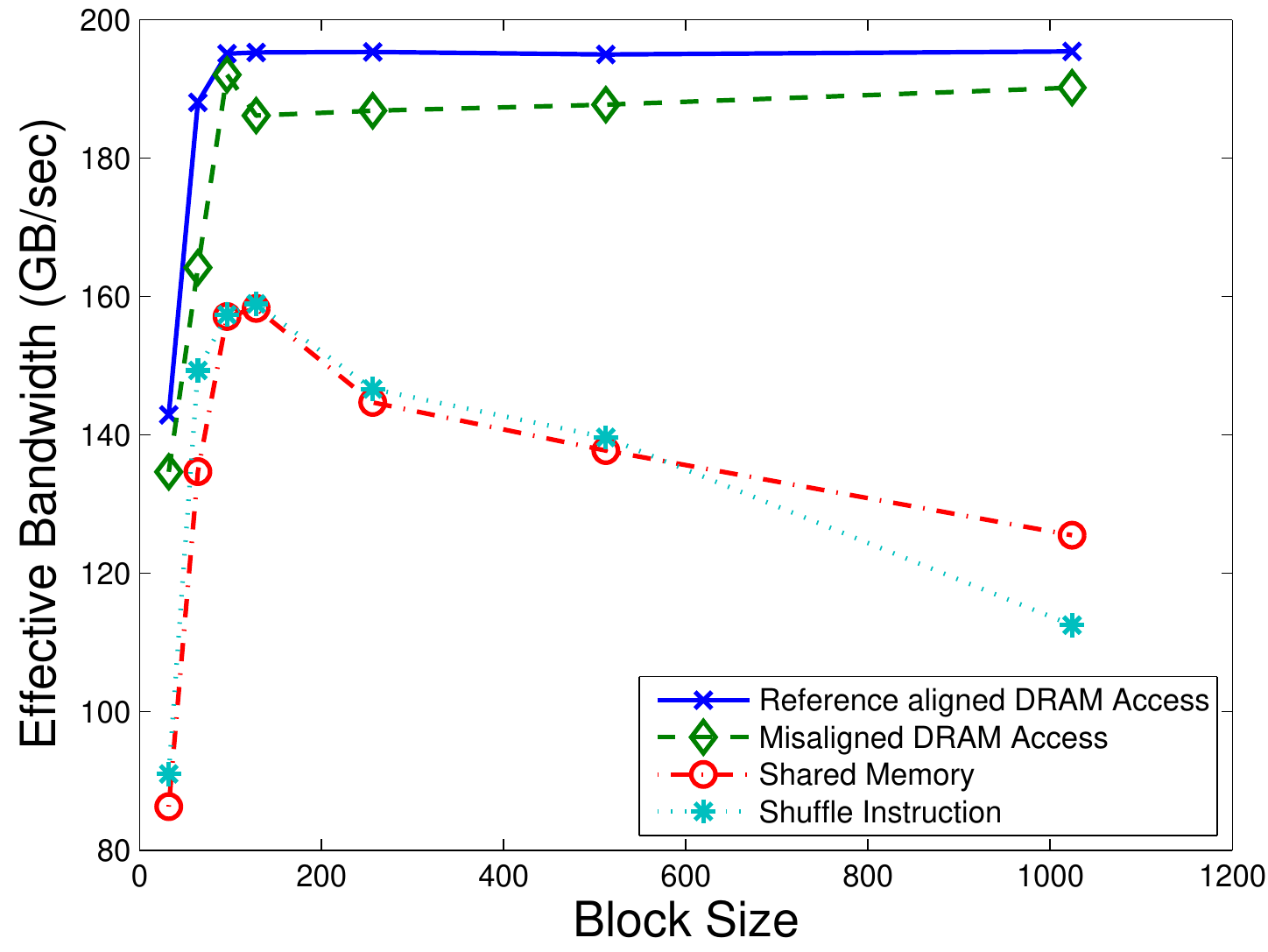}     
\vspace{-10pt}
\caption{Effective bandwidth on K20c for offest-by-one DRAM reads}\label{fig:StreamTimesK20}
\vspace{-10pt}
\end{minipage}

\end{figure*}

Current work on LBM solvers for GPUs either accepts this extra memory transaction (see, for example \cite{Obrecht2013}), or utilises shared memory to perform propagation in the \textit{x} direction, either performing global memory accesses to propagate values between blocks or matching the \textit{x} dimension of the domain to the size of a block \cite{Rinaldi2012, Astorino2011}.

New instructions present within Kepler architecture GPUs allow for propagation within a warp without the use of shared memory, and only require shared memory to propagate values between warps (See Figure \ref{fig:shuffleProp}). In order to determine the  most efficient method for use within an LBM solver, the performance tests in Section \ref{sec:lbmAlg} are repeated for misaligned reading using the three different methods. Figures \ref{fig:StreamTimesK5000} and \ref{fig:StreamTimesK20} present this comparison for both hardware derivatives, in which aligned reads from DRAM are also shown to provide a reference point, as they represent optimal memory behaviour. For both GPUs simple misalignment in DRAM is observed to be the more efficient method, achieving at least a 7.6\% improvement over shared memory and 4.7\% improvement over shuffle memory on a K5000m, rising to 17.6\% and 17.1\% on a K20c. The extra synchronization and intermediate registers required for the use of shared memory or the \textit{shuffle} instruction lowers the achieved bandwidth. Still, for reasonable block sizes (block sizes much above 256 would be impossible to implement in a full LBM code due to the number of registers consumed) the \textit{shuffle} instruction slightly outperforms the use of shared memory alone.

\section{Validation of the LBM code}
To validate the LBM solver, a lid driven cavity case was performed at three Reynolds numbers \{100, 400, 1000\} and compared against reference data from \cite{Jiang1994}. In each case a cubic domain is created with a non-zero \textit{x} velocity on the top wall and non-slip conditions on every other wall. The boundary conditions from \cite{Hecht2010} are used to create the stationary and moving boundaries, and scaling is controlled according to

\begin{equation}\label{eqn:LMScaling}
u=\frac{(\tau -0.5)Re}{3L}
\end{equation}

\begin{figure}[ht!]
		\begin{center}
		\begin{subfigure}[b]{0.47\textwidth}	
			 \includegraphics[width=\textwidth]{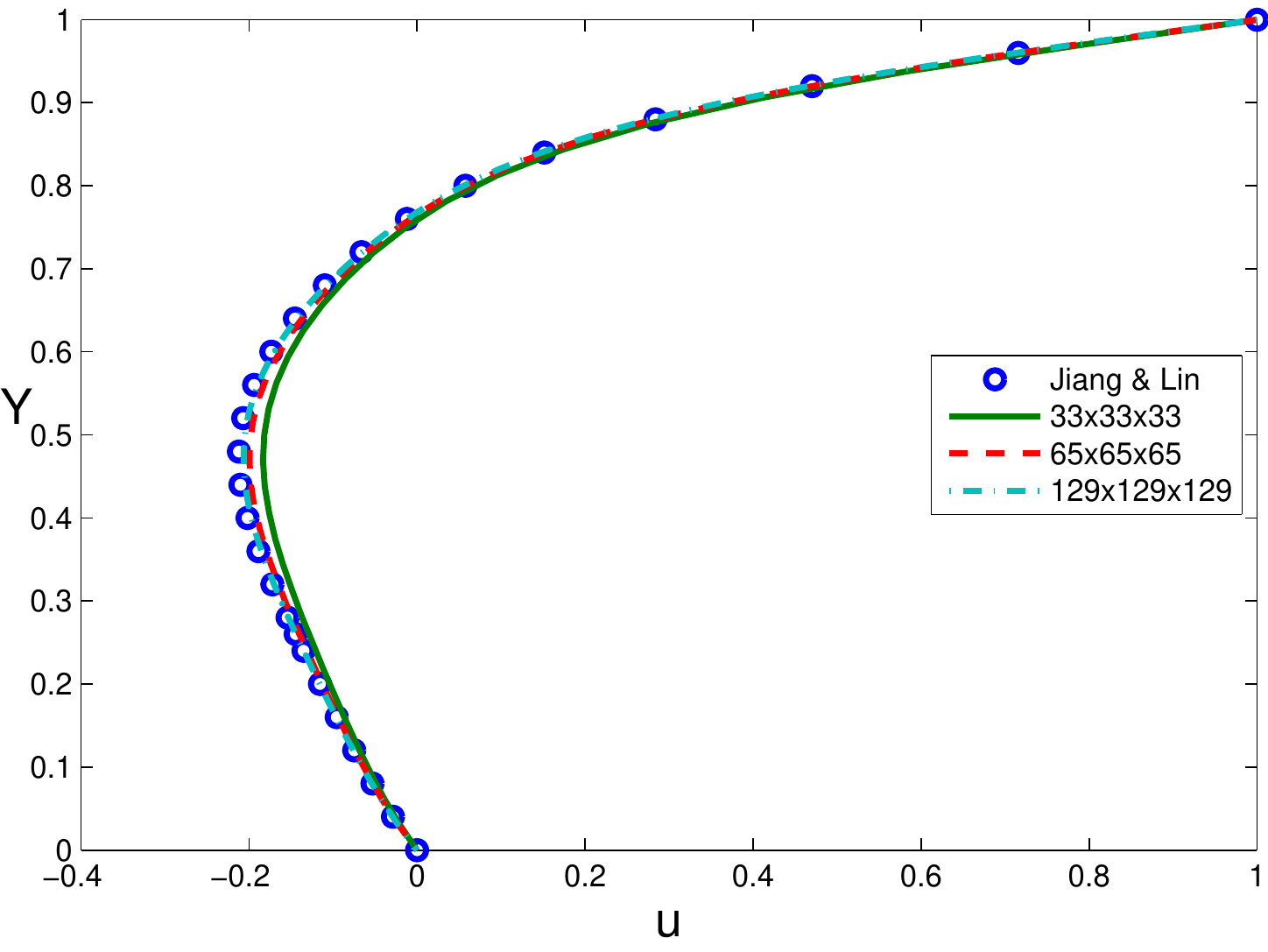}   
	 		\caption{Re=100}\label{fig:3DRe=100}  
		\end{subfigure}
		\begin{subfigure}[b]{0.47\textwidth}	
			 \includegraphics[width=\textwidth]{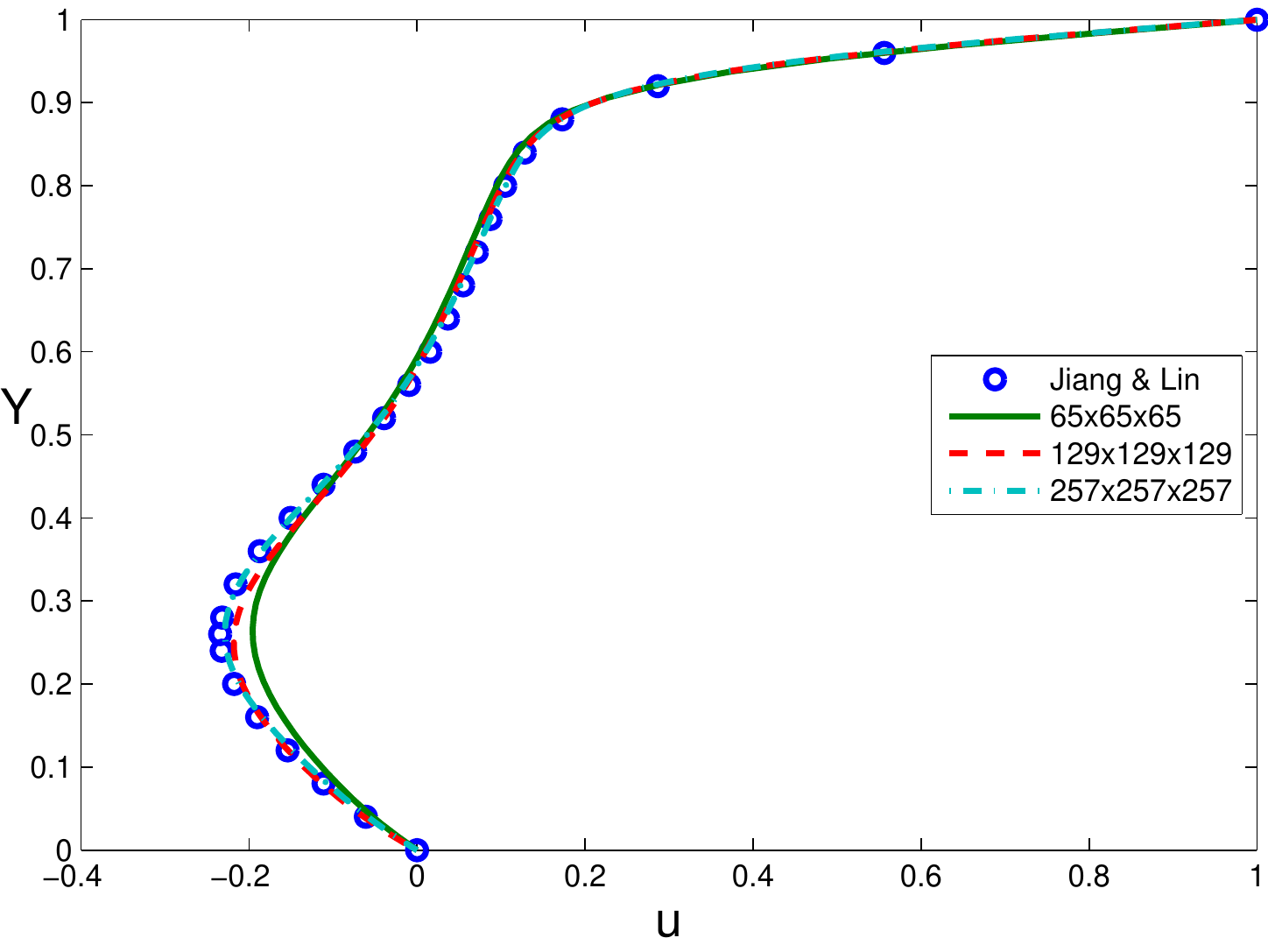}   
			 \caption{Re=400}\label{fig:3DRe=400}  
		\end{subfigure}
		\begin{subfigure}[b]{0.47\textwidth}	
			 \includegraphics[width=\textwidth]{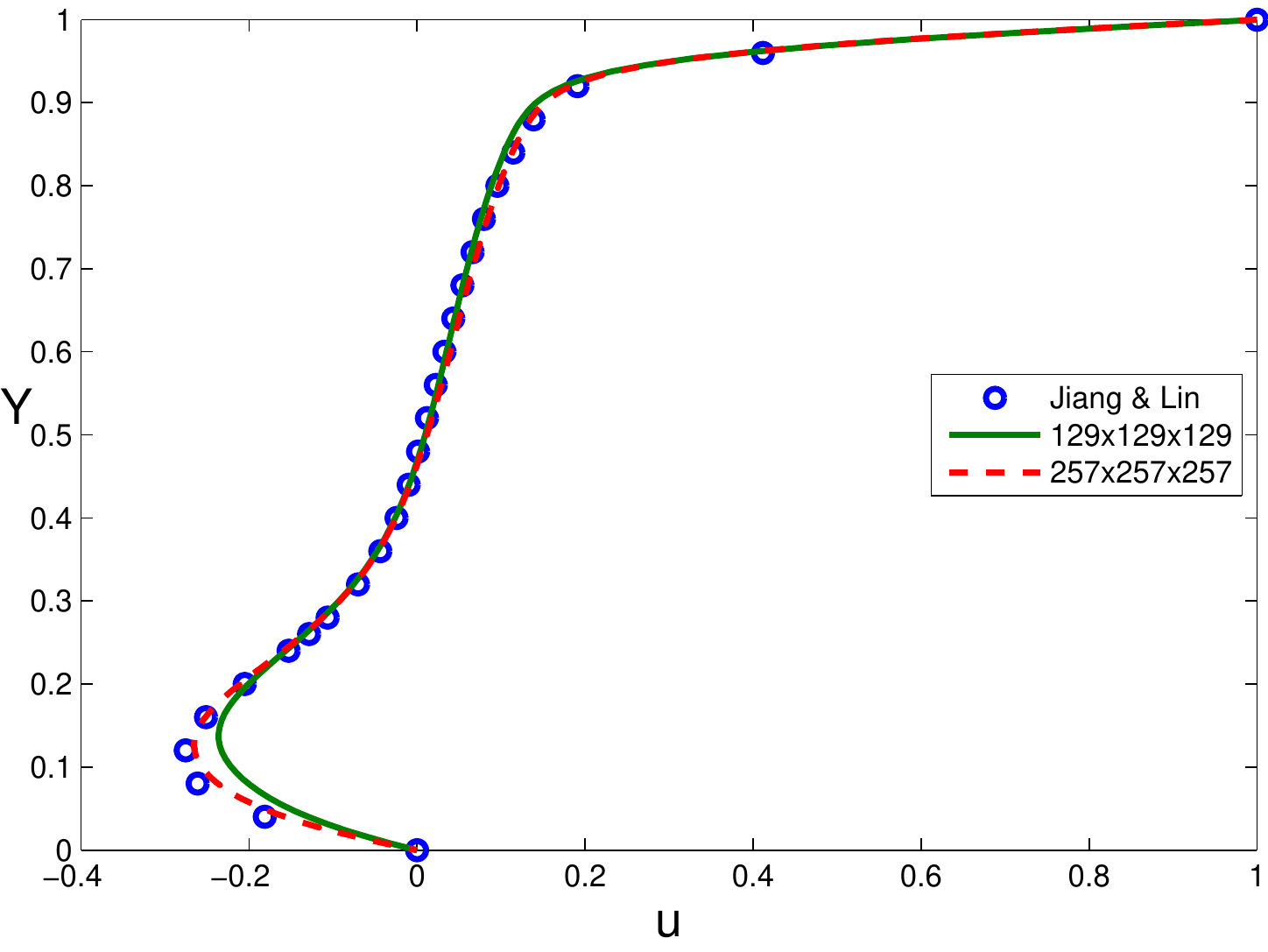}   
	 		\caption{Re=1000}\label{fig:3DRe=1000}  
	\end{subfigure}	
	\end{center}
		\vspace{-10pt}
	\caption{Lid Driven Cavity Validation}\label{fig:LDCConvergence}
		\vspace{-10pt}
	\end{figure}

Figure \ref{fig:LDCConvergence} displays profiles of velocity extracted from the centre of the 3D domain and the resolutions used. For higher Reynolds number computations the domain sizes are increased to reflect the higher levels of resolution required. In all cases the results demonstrate a convergence for subsequent increase in lattice size and reference results are shown to be reproduced.

\section{Performance}
One hundred iterations of the lid driven cavity test case were performed over a variety of block sizes for domains up to a size of $256^2$ in single precision on K5000m and K20c GPUs, with the domain size limited by DRAM size. The blocks are kept as $x$ dimension dominant as possible to facilitate improved cache use, with the obvious exception that threads must not be allocated to indices outside of the computational domain. In this event multiple rows of the highest common denominator between the domain and the desired block size are used. Tables \ref{table:3DLBMPerfK5000} and \ref{table:3DLBMPerfK20} show the mean performance and standard deviation of the 100 iterations in MLUPS. Peak performance is 420 MLUPS on the K5000m and 1036 MLUPS on the K20c.

\begin{figure*}

\begin{minipage}{0.49\textwidth}
\includegraphics[width=\linewidth]{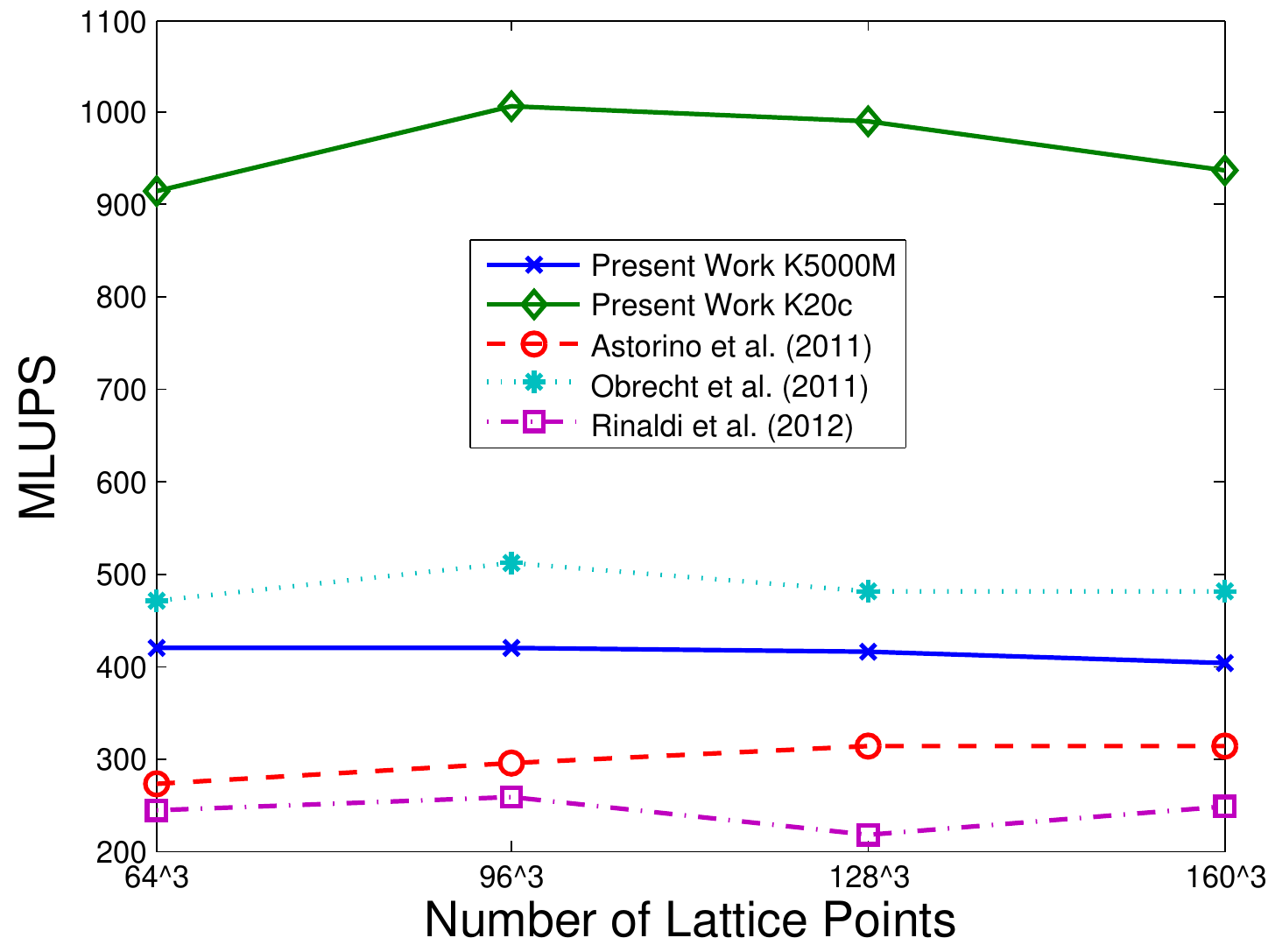}     
\vspace{-10pt}
\caption{Absolute performance}\label{fig:AbsLBMPerf}
\vspace{-10pt}
\end{minipage}
\hspace{0.5cm}
\begin{minipage}{0.49\textwidth}
\includegraphics[width=\linewidth]{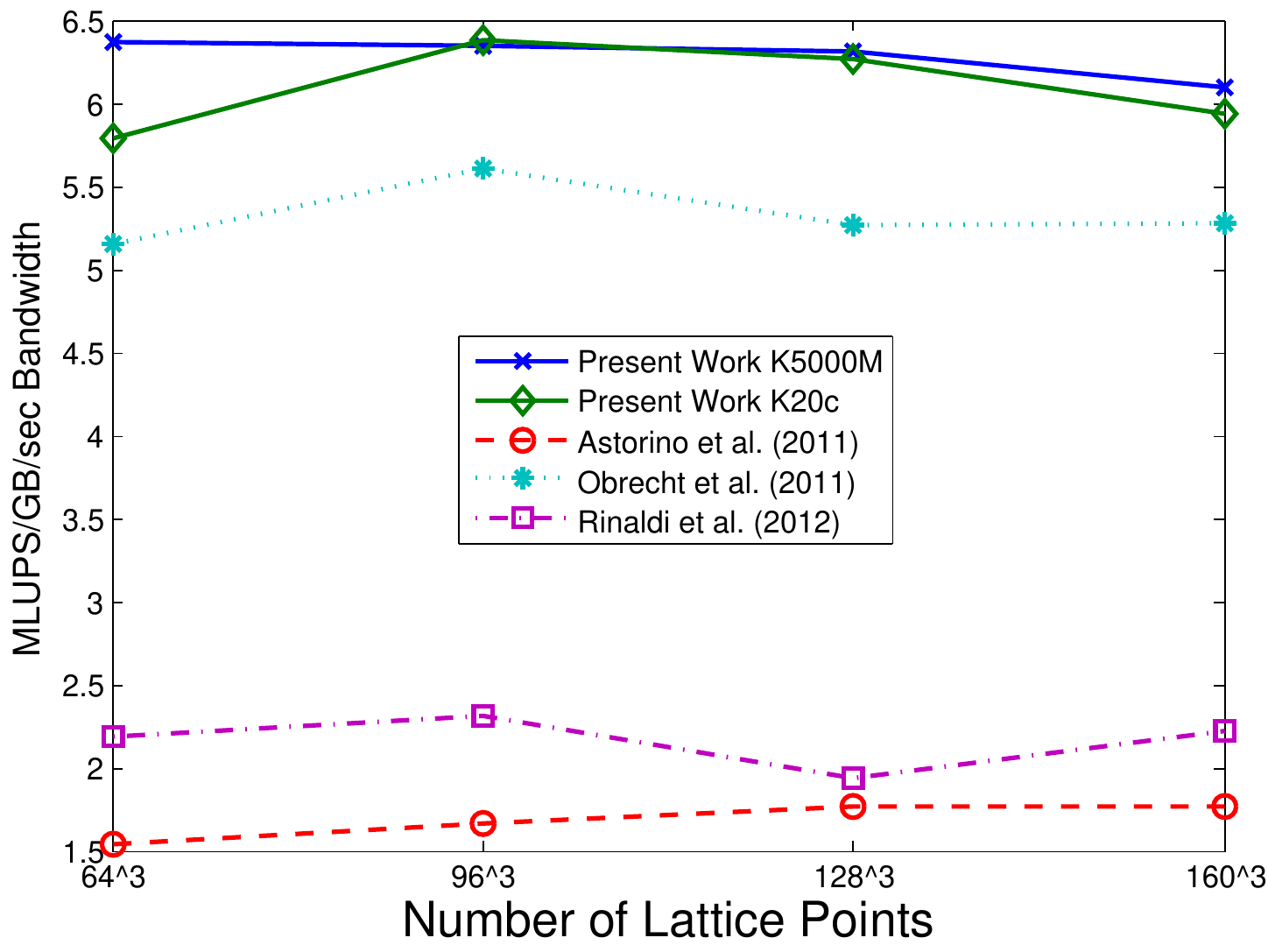}     
\vspace{-10pt}
\caption{Performance scaled for bandwidth}\label{fig:ScaledLBMPerf}
\vspace{-10pt}
\end{minipage}

\end{figure*}

Figure \ref{fig:AbsLBMPerf} displays the performance of the present LBM solver compared to implementations on previous hardware generations found in work by Obrecht et al.\cite{Obrecht2011a} (compute 1.3), Rinaldi et al.\cite{Rinaldi2012} (compute 1.3) and Astorino et al.\cite{Astorino2011} (compute 2.0), which are also tabulated in Table \ref{table:3DMLUPS}. All results are reported for single precision calculations and include the calculation of boundary condition values.

Figure \ref{fig:ScaledLBMPerf} scales the performance relative to the measured bandwidth of the device (theoretical bandwidth is used for Astorino et al. as a measured bandwidth is not reported) and it can be seen that the present implementation is as efficient (slightly more in the case of the K5000M) than the current best implementation found elsewhere in literature. It is worth noting that, following the examination of the use of shared memory in section \ref{sec:StreamAccess}, the implementations found in Astorino et al \cite{Astorino2011} and Rinaldi et al \cite{Rinaldi2012} both make use of shared memory.

%\FloatBarrier

\begin{figure}
\begin{center}
\includegraphics[width=\linewidth]{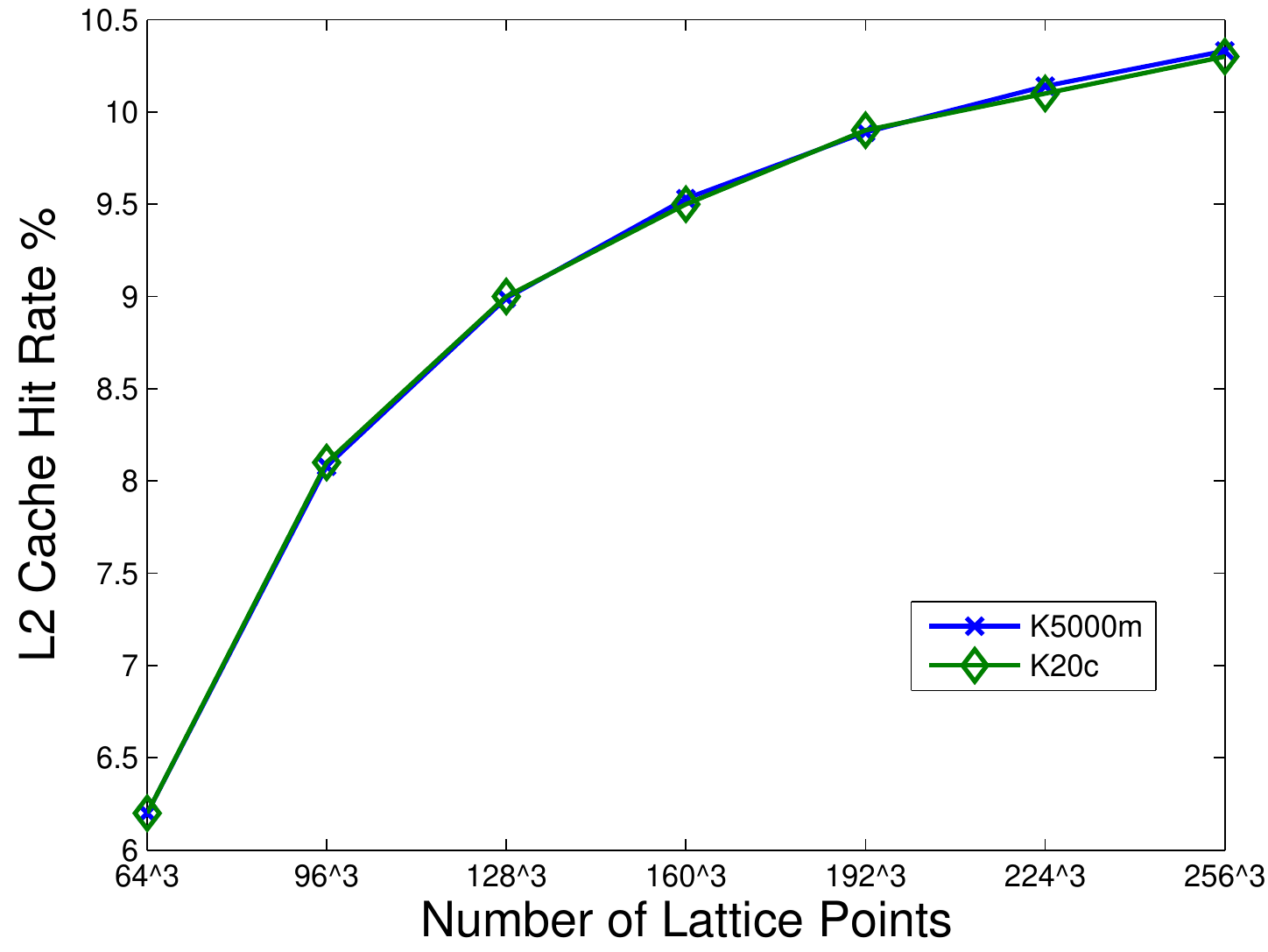}     
\end{center}
\vspace{-10pt}
\caption{L2 cache hit rates}\label{fig:LBML2Cache}
\vspace{-10pt}
\end{figure}

The L2 cache is used exclusively within the LBM solver when misaligned accesses in the stream operation occur, every other operation is fully coalesced and has no re-use of fetched data. The hit rate for L2 cache will therefore depend on the ratio of misaligned to aligned accesses in the stream operation, which will be constant, except for misaligned accesses that would fall outside the boundaries of the domain, and are therefore not performed. This non-constant reduction in misaligned accesses (and therefore L2 cache use) is proportional to the ratio of boundary points to interior domain points, and will tend towards zero (and therefore constant L2 cache use) as the $x$ dimension increases and the ratio of misaligned to aligned accesses decreases, as shown  in Figure \ref{fig:LBML2Cache}.

%		\begin{figure}
%		\begin{center}
%	 \includegraphics[width=0.7\textwidth]{Fig17.pdf}     
%		\end{center}
%	\vspace{-10pt}
%		\caption{Performance results on domain sizes up to $256^3$}\label{fig:LBMK20K5000}
%	\vspace{-10pt}
%	\end{figure}
%	\FloatBarrier
%	\begin{center}

Analysing the performance results of solver reveals interesting behaviour in the variation of performance. One would expect truly random variation to manifest itself in the form of a Gaussian-like distribution centred about the mean value. Whilst, at smaller domain sizes this is true, the vast majority of results exhibit periodic performance variation. A fourier analysis of the results was performed to ascertain the amount of periodic contribution to performance variation, and the frequency at which it occurs. Below sample results for a domain of size $192^3$ are presented.

\begin{figure}
\begin{center}
\includegraphics[width=\linewidth]{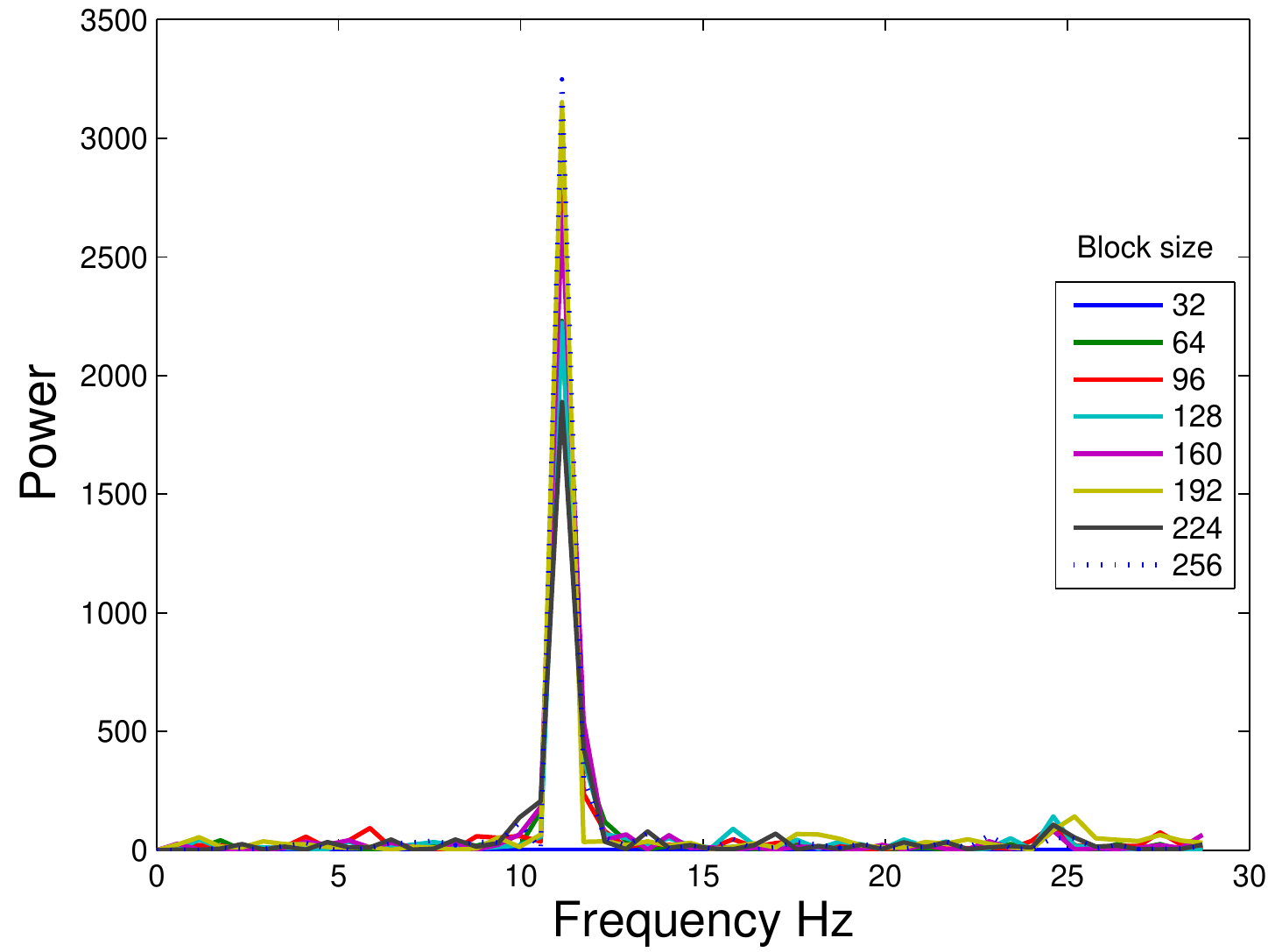}     
\end{center}
\vspace{-10pt}
\caption{Fourier analysis - domain size = $192^3$}\label{fig:3d192Fft}
\vspace{-10pt}
\end{figure}

As figure \ref{fig:3d192Fft} shows, there is clearly a frequency component to the variation in performance. In this case the maximum contribution is found at a frequency of 11.71Hz. Performing the same analysis across all of the test cases yields the following results:

No concrete explanation for this variation in performance can be provided, as it is likely a hardware level issue. Likely candidates are power and/or thermal management strategies of the GPU. The low-level analysis required to determine the cause of this variation is beyond the scope of the present work, although further study of this behaviour should be conducted as it has implications for all GPU programming.

\section{Conclusion}
This work has demonstrated the optimization and validation of a 3D GPU-based Lattice Boltzmann solver on Kepler architecture GPUs. The use of shared memory, and an intrinsic memory-less intra-warp \textit{shuffle} operation, have been shown to be ineffective at improving the performance of the memory-intensive streaming operation, in spite of the fact that their use increases the number of coalesced accesses to DRAM. Instead, a `naive' implementation that has misaligned access to DRAM is found to be more effective due to its lower register usage and no need for any additional control flow. Using this information an efficient Lattice Boltzmann solver was programmed and benchmarked on GK104 and GK110 generation GPUs, achieving a peak performance of up to 1036 MLUPS on GK110 GPUs. The findings of this work have already been applied in the design of an interactive two dimensional LBM solver, where the high performance of the fluid solver allows boundary conditions to be captured from real-world geometry using an infrared depth sensor while maintaining real-time fluid flow evolution and visualization \cite{Mawson2013}.

\begin{table*}
\begin{center}

\begin{tabular}{c | c c c c c c c}
\multicolumn{8}{c}{\textbf{Domain Size}} \\ \hline 
\textbf{Block Size} & \textbf{$64^3$} & \textbf{$96^3$} & \textbf{$128^3$} & \textbf{$160^3$} & \textbf{$192^3$} & \textbf{$224^3$} & \textbf{$256^3$} \\ \hline \hline
\multirow{2}{*}{\textbf{32}} & 321.74 & 331.07 & 344.9 & 343.37 & 346.63 & 348.17 & 358.02\\
 & 0.64 & 0.21 & 0.1 & 0.05 & 0.03 & 0.08 & 0.03\\ \hline
\multirow{2}{*}{\textbf{64}} & 419.56 & 416.49 & 413.47 & 401.63 & 413.29 & 392.25 & 412.17\\
 & 2.64 & 1.31 & 1.3 & 0.91 & 0.85 & 0.72 & 0.67\\ \hline
\multirow{2}{*}{\textbf{96}} & 416.09 & 418.21 & 411.79 & 400.82 & 413.84 & 391.49 & 409.74\\
 & 2.29 & 1.36 & 1.18 & 0.94 & 0.94 & 0.72 & 0.66\\ \hline
\multirow{2}{*}{\textbf{128}} & \textbf{\textit{419.93}} & 416.07 & 415.15 & 400.76 & 413.51 & 391.73 & 413.01\\
 & 2.49 & 1.47 & 1.37 & 0.88 & 0.86 & 0.71 & 0.7\\ \hline
\multirow{2}{*}{\textbf{160}} & 416.18 & 416 & 412.04 & \textbf{\textit{402.63}} & 411.97 & 391.68 & 409.59\\
 & 2.38 & 1.31 & 1.17 & 0.89 & 0.93 & 0.71 & 0.61\\ \hline
\multirow{2}{*}{\textbf{192}} & 420.38 & \textbf{\textit{418.89}} & 413.81 & 400.33 & \textbf{\textit{415.62}} & 392.66 & 410.82\\
 & 3.17 & 1.35 & 1.46 & 0.89 & 0.96 & 0.71 & 0.7\\ \hline
\multirow{2}{*}{\textbf{224}} & 415.23 & 415.96 & 412.41 & 400.68 & 412.06 & 390.92 & 409.93\\
 &2.62 & 1.3 & 1.29 & 0.89 & 0.84 & 0.71 & 0.64\\ \hline
\multirow{2}{*}{\textbf{256}} & 420.36 & 416.6 & \textbf{\textit{416.27}} & 400.75 & 414.52 & \textbf{\textit{394.93}} & \textbf{\textit{414.47}}\\
 & 2.37 & 1.28 & 1.57 & 0.93 & 0.96 & 0.74 & 0.7\\ \hline

\end{tabular}%
\caption{Mean performance and $\sigma$ in MLUPS for the 3D LBM solver on K5000m}\label{table:3DLBMPerfK5000}
\end{center}
\end{table*}

%------------------

\begin{table*}
\begin{center}
%\resizebox{\textwidth}{!}{%
\begin{tabular}{c | c c c c c c c}

 \multicolumn{8}{c}{\textbf{Domain Size}} \\ \hline 
\textbf{Block Size} & \textbf{$64^3$} & \textbf{$96^3$} & \textbf{$128^3$} & \textbf{$160^3$} & \textbf{$192^3$} & \textbf{$224^3$} & \textbf{$256^3$} \\ \hline \hline
\multirow{2}{*}{\textbf{32}} & 674.75 & 701.66 & 725.22 & 714.54 & 723.45 & 722.35 & 739.12\\
 & 1.32 & 0.61 & 1.21 & 0.23 & 0.3 & 0.14 & 0.11\\ \hline
\multirow{2}{*}{\textbf{64}} & 909.79 & 1006.73 & 979.22 & \textbf{\textit{937.35}} & \textbf{\textit{1036.41}} & 944.2 & 991.28\\
 & 3.48 & 2.02 & 1.07 & 1.92 & 1.5 & 1.65 & 0.88\\ \hline
\multirow{2}{*}{\textbf{96}} & 904.91 & 981.86 & 957.39 & 918.37 & 1008.64 & 990.82 & 962.44\\
 & 4.64 & 1.39 & 1.76 & 1.78 & 0.5 & 3.67 & 1.1\\ \hline
\multirow{2}{*}{\textbf{128}} & 912.9 & \textbf{\textit{1007.04}} & 981.79 & 935.4 & 1035.36 & \textbf{\textit{997.7}} & 990.23\\
 & 3.62 & 2.22 & 1.43 & 1.96 & 0. & 1.74 & 1.09\\ \hline
\multirow{2}{*}{\textbf{160}} & \textbf{\textit{913.96}} & 962.88 & 986.46 & 912.16 & 985.47 & 987.15 & 1001.31\\
 & 3.17 & 1.29 & 3.6 & 1.6 & 0.77 & 0.56 & 4.9\\ \hline
\multirow{2}{*}{\textbf{192}} & 905.86 & 957.74 & 974.51 & 911.49 & 977.51 & 972.65 & 996.21\\
 & 2.67 & 0.69 & 0.81 & 2.3 & 0.29 & 0.4 & 0.61\\ \hline
\multirow{2}{*}{\textbf{224}} & 869.78 & 893.35 & 949.74 & 900.33 & 936.42 & 938.65 & 961.36\\
 & 2.24 & 0.75 & 1.1 & 2.66 & 0.31 & 0.44 & 0.64\\ \hline
\multirow{2}{*}{\textbf{256}} & 902.12 & 948.44 & \textbf{\textit{990.39}} & 912.45 & 974.05 & 977.94 & \textbf{\textit{1020.59}}\\
 & 3.78 & 1.2 & 1.22 & 1.63 & 0.41 & 0.31 & 0.7\\ \hline

\end{tabular}%
%}
\caption{Mean performance and $\sigma$ in MLUPS for the 3D LBM solver on K20c}\label{table:3DLBMPerfK20}
\end{center}
\end{table*}

%------------------

\begin{table*}
\begin{center}
%\resizebox{\textwidth}{!}{%
\begin{tabular}{r r r r r r}
\hline
\textbf{Domain Size} & \textbf{K5000m} & \textbf{K20c} & \textbf{Astorino(2011)}& \textbf{Obrecht(2011)}& \textbf{Rinaldi(2012)}\\ \hline \hline
$64^3$ &420 & 914 & 273 & 471 & 273 \\  
$96^3$ & 419 & 1007 & 296 & 512 & 296\\ 
$128^3$ & 416 & 990 & 314 & 481 & 314\\ 
$160^3$ & 403 & 937 & 313 & 482 & 313\\ \hline
\end{tabular}%
%}
\caption{Performance expressed in MLUPS for K5000m and K20c GPUs, compared against existing work}\label{table:3DMLUPS}
\end{center}
\end{table*}

%------------------

\begin{table*}
\begin{center}
%\resizebox{\textwidth}{!}{%
\begin{tabular}{r r r r r r}
\hline
\textbf{Domain Size} & \textbf{K5000m} & \textbf{K20c} & \textbf{Astorino(2011)}& \textbf{Obrecht(2011)}& \textbf{Rinaldi(2012)}\\ \hline \hline
$64^3$ & 6.4714 & 5.7886 & 1.539 & 5.233 & 2.579  \\ 
$96^3$ & 6.4485 & 6.3781 & 1.669 & 5.689 & 2.726 \\ 
$128^3$ & 6.4081 & 6.2727 & 1.77 & 5.344 & 2.284 \\ 
$160^3$ & 5.1981 & 5.9368 & 1.764 & 5.356 & 2.61 \\ \hline
\end{tabular}%
%}
\caption{Performance normalized against DRAM bandwidth (MLUPS/GB/sec)}\label{table:3DMLUPSscaled}
\end{center}
\end{table*}

%------------------

\begin{table*}
\begin{center}
%\resizebox{\textwidth}{!}{%

\begin{tabular}{l | c c c c c c c}
\hline
\textbf{Domain Size }& $64^3$ & $96^3$ & $128^3$ & $160^3$ & $192^3$ & $224^3$ & $256^3$ 
\\ \hline \hline
\textbf{Peak Frequency Component (Hz)} \\
K5000M & 11.28 & 15.98 & 9.41 & 11.87 & 11.71 & 11.14 & 11.36\\
K20c & N/A & 10.72 & 9.445 & 11.15 & 11.01 & 11.31 & 11.56\\ 
\hline
\end{tabular}%

%}
\caption{Frequency components of 3D LBM performance on K5000m}\label{table:LBMFreq3dK5000}
\end{center}
\end{table*}

\label{}



%% References


\bibliographystyle{model1a-num-names}
\bibliography{library}

\begin{thebibliography}{34}
\expandafter\ifx\csname natexlab\endcsname\relax\def\natexlab#1{#1}\fi
\providecommand{\bibinfo}[2]{#2}
\ifx\xfnm\relax \def\xfnm[#1]{\unskip,\space#1}\fi
%Type = Article
\bibitem[{Chen and Doolen(1998)}]{Chen1998}
\bibinfo{author}{S.~Chen}, \bibinfo{author}{G.~Doolen},
  \bibinfo{journal}{Annual review of fluid mechanics} \bibinfo{volume}{30}
  (\bibinfo{year}{1998}) \bibinfo{pages}{329--364}.
%Type = Book
\bibitem[{Succi(2001)}]{Succi2001}
\bibinfo{author}{S.~Succi}, \bibinfo{title}{{The Lattice Boltzmann Equation for
  Fluid Dynamics and Beyond (Numerical Mathematics and Scientific
  Computation)}}, \bibinfo{publisher}{Oxford University Press, USA},
  \bibinfo{year}{2001}.
%Type = Article
\bibitem[{Shan et~al.(2006)Shan, Yuan, and Chen}]{Shan2006}
\bibinfo{author}{X.~Shan}, \bibinfo{author}{X.-F. Yuan},
  \bibinfo{author}{H.~Chen}, \bibinfo{journal}{Journal of Fluid Mechanics}
  \bibinfo{volume}{550} (\bibinfo{year}{2006}) \bibinfo{pages}{413}.
%Type = Article
\bibitem[{Li et~al.(2003)Li, Wei, and Kaufman}]{Li2003}
\bibinfo{author}{W.~Li}, \bibinfo{author}{X.~Wei},
  \bibinfo{author}{A.~Kaufman}, \bibinfo{journal}{The Visual Computer}
  (\bibinfo{year}{2003}).
%Type = Article
\bibitem[{Jan\ss~en and Krafczyk(2011)}]{Janßen2011}
\bibinfo{author}{C.~Jan\ss~en}, \bibinfo{author}{M.~Krafczyk},
  \bibinfo{journal}{Computers \& Mathematics with Applications}
  \bibinfo{volume}{61} (\bibinfo{year}{2011}) \bibinfo{pages}{3549--3563}.
%Type = Article
\bibitem[{Obrecht et~al.(2011)Obrecht, Kuznik, Tourancheau, and
  Roux}]{Obrecht2011}
\bibinfo{author}{C.~Obrecht}, \bibinfo{author}{F.~Kuznik},
  \bibinfo{author}{B.~Tourancheau}, \bibinfo{author}{J.-J. Roux},
  \bibinfo{journal}{Computers \& Fluids} \bibinfo{volume}{54}
  (\bibinfo{year}{2011}) \bibinfo{pages}{118--126}.
%Type = Article
\bibitem[{Miki et~al.(2012)Miki, Wang, Aoki, Imai, Ishikawa, Takase, and
  Yamaguchi}]{Miki2012}
\bibinfo{author}{T.~Miki}, \bibinfo{author}{X.~Wang},
  \bibinfo{author}{T.~Aoki}, \bibinfo{author}{Y.~Imai},
  \bibinfo{author}{T.~Ishikawa}, \bibinfo{author}{K.~Takase},
  \bibinfo{author}{T.~Yamaguchi}, \bibinfo{journal}{Computer methods in
  biomechanics and biomedical engineering} \bibinfo{volume}{15}
  (\bibinfo{year}{2012}) \bibinfo{pages}{771--8}.
%Type = Article
\bibitem[{Ryoo et~al.(2008)Ryoo, Rodrigues, Baghsorkhi, Stone, Kirk, and
  Hwu}]{Ryoo2008}
\bibinfo{author}{S.~Ryoo}, \bibinfo{author}{C.~I. Rodrigues},
  \bibinfo{author}{S.~S. Baghsorkhi}, \bibinfo{author}{S.~S. Stone},
  \bibinfo{author}{D.~B. Kirk}, \bibinfo{author}{W.-m.~W. Hwu},
  \bibinfo{journal}{Proceedings of the 13th ACM SIGPLAN Symposium on Principles
  and practice of parallel programming - PPoPP '08}  (\bibinfo{year}{2008})
  \bibinfo{pages}{73}.
%Type = Article
\bibitem[{Henning(2006)}]{Henning2006}
\bibinfo{author}{J.~Henning}, \bibinfo{journal}{ACM SIGARCH Computer
  Architecture News}  (\bibinfo{year}{2006}).
%Type = Article
\bibitem[{T\"{o}lke and Krafczyk(2008)}]{Tolke2008}
\bibinfo{author}{J.~T\"{o}lke}, \bibinfo{author}{M.~Krafczyk},
  \bibinfo{journal}{International Journal of Computational Fluid Dynamics}
  \bibinfo{volume}{22} (\bibinfo{year}{2008}) \bibinfo{pages}{443--456}.
%Type = Article
\bibitem[{T\"{o}lke(2008)}]{Tolke2008a}
\bibinfo{author}{J.~T\"{o}lke}, \bibinfo{journal}{Computing and Visualization
  in Science} \bibinfo{volume}{13} (\bibinfo{year}{2008})
  \bibinfo{pages}{29--39}.
%Type = Article
\bibitem[{Habich et~al.(2011)Habich, Zeiser, Hager, and Wellein}]{Habich2011}
\bibinfo{author}{J.~Habich}, \bibinfo{author}{T.~Zeiser},
  \bibinfo{author}{G.~Hager}, \bibinfo{author}{G.~Wellein},
  \bibinfo{journal}{Advances in Engineering Software} \bibinfo{volume}{42}
  (\bibinfo{year}{2011}) \bibinfo{pages}{266--272}.
%Type = Inproceedings
\bibitem[{Obrecht et~al.(2011{\natexlab{a}})Obrecht, Kuznik, Tourancheau, and
  Roux}]{Obrecht2011c}
\bibinfo{author}{C.~Obrecht}, \bibinfo{author}{F.~Kuznik},
  \bibinfo{author}{B.~Tourancheau}, \bibinfo{author}{J.-J. Roux}, in:
  \bibinfo{booktitle}{Proceedings of the 9th international conference on High
  performance computing for computational science},
  \bibinfo{publisher}{Springer-Verlag}, \bibinfo{address}{Berkeley, CA},
  \bibinfo{year}{2011}{\natexlab{a}}, pp. \bibinfo{pages}{151--161}.
%Type = Article
\bibitem[{Obrecht et~al.(2011{\natexlab{b}})Obrecht, Kuznik, Tourancheau, and
  Roux}]{Obrecht2011a}
\bibinfo{author}{C.~Obrecht}, \bibinfo{author}{F.~Kuznik},
  \bibinfo{author}{B.~Tourancheau}, \bibinfo{author}{J.-J. Roux},
  \bibinfo{journal}{Computers \& Mathematics with Applications}
  \bibinfo{volume}{61} (\bibinfo{year}{2011}{\natexlab{b}})
  \bibinfo{pages}{3628--3638}.
%Type = Article
\bibitem[{Favier et~al.(2013)Favier, Revell, and Pinelli}]{Favier2013}
\bibinfo{author}{J.~Favier}, \bibinfo{author}{A.~Revell},
  \bibinfo{author}{A.~Pinelli}  (\bibinfo{year}{2013}).
%Type = Inproceedings
\bibitem[{Mawson et~al.(2013)Mawson, Leaver, and Revell}]{Mawson2013}
\bibinfo{author}{M.~Mawson}, \bibinfo{author}{G.~Leaver},
  \bibinfo{author}{A.~Revell}, in: \bibinfo{booktitle}{NAFEMS World Congress
  2013 Summary of Proceedings}, \bibinfo{publisher}{NAFEMS Ltd},
  \bibinfo{address}{Salzburg}, \bibinfo{year}{2013}, p. \bibinfo{pages}{204}.
%Type = Article
\bibitem[{Obrecht et~al.(2013{\natexlab{a}})Obrecht, Kuznik, Tourancheau, and
  Roux}]{Obrecht2013}
\bibinfo{author}{C.~Obrecht}, \bibinfo{author}{F.~Kuznik},
  \bibinfo{author}{B.~Tourancheau}, \bibinfo{author}{J.-J. Roux},
  \bibinfo{journal}{Computers \& Mathematics with Applications}
  \bibinfo{volume}{65} (\bibinfo{year}{2013}{\natexlab{a}})
  \bibinfo{pages}{252--261}.
%Type = Article
\bibitem[{Obrecht et~al.(2013{\natexlab{b}})Obrecht, Kuznik, Tourancheau, and
  Roux}]{obrecht2013b}
\bibinfo{author}{C.~Obrecht}, \bibinfo{author}{F.~Kuznik},
  \bibinfo{author}{B.~Tourancheau}, \bibinfo{author}{J.-J. Roux},
  \bibinfo{journal}{Computers \& Fluids} \bibinfo{volume}{80}
  (\bibinfo{year}{2013}{\natexlab{b}}) \bibinfo{pages}{269--275}.
%Type = Article
\bibitem[{Habich et~al.(2012)Habich, Feichtinger, K\"{o}stler, Hager, and
  Wellein}]{Habich2012}
\bibinfo{author}{J.~Habich}, \bibinfo{author}{C.~Feichtinger},
  \bibinfo{author}{H.~K\"{o}stler}, \bibinfo{author}{G.~Hager},
  \bibinfo{author}{G.~Wellein}, \bibinfo{journal}{Computers \& Fluids}
  (\bibinfo{year}{2012}) \bibinfo{pages}{1--7}.
%Type = Unpublished
\bibitem[{{NVIDIA Corporation}(2012)}]{NVIDIACorporation2012}
\bibinfo{author}{{NVIDIA Corporation}}, \bibinfo{title}{{NVIDIA’s Next
  Generation CUDA Compute Architecture: Kepler GK110. The Fastest, Most
  Efficient HPC Architecture Ever Built}}, \bibinfo{year}{2012}.
%Type = Book
\bibitem[{NVIDIA(2009)}]{NVIDIA2009a}
\bibinfo{author}{NVIDIA}, \bibinfo{title}{{CUDA Programming Guide Version
  2.3.1}}, \bibinfo{year}{2009}.
%Type = Techreport
\bibitem[{NVIDIA(2010)}]{NVIDIAPG3.2}
\bibinfo{author}{NVIDIA}, \bibinfo{title}{{NVIDIA CUDA C Programming Guide
  Version 3.2}}, \bibinfo{type}{Technical Report}, NVIDIA Corporation,
  \bibinfo{year}{2010}.
%Type = Article
\bibitem[{Luebke et~al.(2007)Luebke, Humphreys, and Res}]{Luebke2007}
\bibinfo{author}{D.~Luebke}, \bibinfo{author}{G.~Humphreys},
  \bibinfo{author}{N.~Res}, \bibinfo{journal}{Computer} \bibinfo{volume}{40}
  (\bibinfo{year}{2007}) \bibinfo{pages}{96--100}.
%Type = Book
\bibitem[{Pharr and Fernando(2005)}]{Pharr2005}
\bibinfo{author}{M.~Pharr}, \bibinfo{author}{R.~Fernando}, \bibinfo{title}{{GPU
  Gems 2: Programming Techniques for High-Performance Graphics and
  General-Purpose Computation}}, \bibinfo{publisher}{Addison-Wesley},
  \bibinfo{year}{2005}.
%Type = Article
\bibitem[{Bhatnagar and Gross(1954)}]{Bhatnagar1954}
\bibinfo{author}{P.~Bhatnagar}, \bibinfo{author}{E.~Gross},
  \bibinfo{journal}{Physical Review}  (\bibinfo{year}{1954}).
%Type = Article
\bibitem[{He and Luo(1997)}]{He1997b}
\bibinfo{author}{X.~He}, \bibinfo{author}{L.-S. Luo}, \bibinfo{journal}{Journal
  of Statistical Physics} \bibinfo{volume}{88} (\bibinfo{year}{1997})
  \bibinfo{pages}{927--944}.
%Type = Article
\bibitem[{Qian et~al.(1992)Qian, D'Humi\`{e}res, and Lallemand}]{Qian1992}
\bibinfo{author}{Y.~H. Qian}, \bibinfo{author}{D.~D'Humi\`{e}res},
  \bibinfo{author}{P.~Lallemand}, \bibinfo{journal}{Europhysics Letters (EPL)}
  \bibinfo{volume}{17} (\bibinfo{year}{1992}) \bibinfo{pages}{479--484}.
%Type = Article
\bibitem[{Pohl et~al.(2003)Pohl, Kowarschik, and Wilke}]{Pohl2003a}
\bibinfo{author}{T.~Pohl}, \bibinfo{author}{M.~Kowarschik},
  \bibinfo{author}{J.~Wilke}, \bibinfo{journal}{Parallel Processing \ldots}
  \bibinfo{volume}{10} (\bibinfo{year}{2003}).
%Type = Article
\bibitem[{Wellein et~al.(2006)Wellein, Zeiser, Hager, and Donath}]{Wellein2006}
\bibinfo{author}{G.~Wellein}, \bibinfo{author}{T.~Zeiser},
  \bibinfo{author}{G.~Hager}, \bibinfo{author}{S.~Donath},
  \bibinfo{journal}{Computers \& Fluids} \bibinfo{volume}{35}
  (\bibinfo{year}{2006}) \bibinfo{pages}{910--919}.
%Type = Article
\bibitem[{Volkov(2010)}]{Volkov2010}
\bibinfo{author}{V.~Volkov}, \bibinfo{journal}{Proceedings of the GPU
  Technology Conference, \ldots}  (\bibinfo{year}{2010}).
%Type = Article
\bibitem[{Rinaldi et~al.(2012)Rinaldi, Dari, V\'{e}nere, and
  Clausse}]{Rinaldi2012}
\bibinfo{author}{P.~Rinaldi}, \bibinfo{author}{E.~Dari},
  \bibinfo{author}{M.~V\'{e}nere}, \bibinfo{author}{a.~Clausse},
  \bibinfo{journal}{Simulation Modelling Practice and Theory}
  \bibinfo{volume}{25} (\bibinfo{year}{2012}) \bibinfo{pages}{163--171}.
%Type = Techreport
\bibitem[{Astorino et~al.(2011)Astorino, Sagredo, and
  Quarteroni}]{Astorino2011}
\bibinfo{author}{M.~Astorino}, \bibinfo{author}{J.~Sagredo},
  \bibinfo{author}{A.~Quarteroni}, \bibinfo{title}{{A modular lattice Boltzmann
  solver for GPU computing processors}}, \bibinfo{type}{Technical Report}
  \bibinfo{number}{06}, Mathematics Institute of Computational Science and
  Engineering, \bibinfo{address}{Lucerne, Switzerland}, \bibinfo{year}{2011}.
%Type = Article
\bibitem[{Jiang and Lin(1994)}]{Jiang1994}
\bibinfo{author}{B.-n. Jiang}, \bibinfo{author}{T.~Lin},
  \bibinfo{journal}{Computer Methods in Applied Mechanics}
  \bibinfo{volume}{114} (\bibinfo{year}{1994}) \bibinfo{pages}{213--231}.
%Type = Article
\bibitem[{Hecht and Harting(2010)}]{Hecht2010}
\bibinfo{author}{M.~Hecht}, \bibinfo{author}{J.~Harting},
  \bibinfo{journal}{Journal of Statistical Mechanics: Theory and Experiment}
  \bibinfo{volume}{2010} (\bibinfo{year}{2010}) \bibinfo{pages}{P01018}.

\end{thebibliography}






% \end{thebibliography}

\end{document}